\documentclass[journal,comsoc]{IEEEtran}

\usepackage{cite}
\usepackage{amsmath,amssymb,amsfonts,amsthm}
\usepackage{algorithm,algpseudocode}
\usepackage{graphicx}
\usepackage{textcomp}
\usepackage{xcolor}
\usepackage{booktabs}
\usepackage{multirow}
\usepackage{hyperref}
\usepackage{url}
\usepackage{listings}
\usepackage{tabularx}
\usepackage{subcaption}
\usepackage{balance}
\usepackage[most]{tcolorbox}
\usepackage[table]{xcolor}

\theoremstyle{definition}
\newtheorem{definition}{Definition}

\newtheorem{proposition}{Proposition}

\definecolor{capStrong}{HTML}{FDEAEA}   % strong impact
\definecolor{capMedium}{HTML}{FFF4E5}   % medium impact
\definecolor{capWeak}{HTML}{F7FAFF}     % weak impact
\definecolor{rowlabel}{HTML}{F6F6F6}

\newcolumntype{Y}{>{\centering\arraybackslash}X}
\newtcolorbox{takeaway}[1][]{%
  enhanced,
  colback=blue!3!white,
  colframe=blue!40!black,
  fonttitle=\bfseries,
  coltitle=white,
  colbacktitle=blue!40!black,
  attach boxed title to top left={yshift=-2mm, xshift=4mm},
  boxed title style={sharp corners, size=small},
  sharp corners,
  left=6pt, right=6pt, top=6pt, bottom=4pt,
  title={#1}
}

\hypersetup{
    colorlinks=true,
    linkcolor=blue!60!black,
    citecolor=blue!60!black,
    urlcolor=blue!60!black,
}

\newenvironment{packeditemize}{
	\begin{list}{$\bullet$}{
			\setlength{\labelwidth}{4pt}
			\setlength{\itemsep}{0pt}
			\setlength{\leftmargin}{\labelwidth}
			\addtolength{\leftmargin}{\labelsep}
			\setlength{\parindent}{0pt}
			\setlength{\listparindent}{\parindent}
			\setlength{\parsep}{0pt}
			\setlength{\topsep}{1pt}}}{\end{list}}

\lstdefinelanguage{Solidity}{
    keywords={pragma, solidity, contract, interface, function, returns, return, event, emit, mapping, uint256, address, bool, modifier, require, override, virtual, public, external, view, internal, memory, storage, if, else, struct},
    keywordstyle=\color{blue!80!black}\bfseries,
    sensitive=true,
    comment=[l]{//},
    morecomment=[s]{/*}{*/},
    commentstyle=\color{green!50!black}\itshape,
    stringstyle=\color{red!60!black},
    morestring=[b]",
    morestring=[b]',
}

\def\BibTeX{{\rm B\kern-.05em{\sc i\kern-.025em b}\kern-.08em
    T\kern-.1667em\lower.7ex\hbox{E}\kern-.125emX}}

\begin{document}

% ══════════════════════════════════════════════════════════════════════════════
%  TITLE
% ══════════════════════════════════════════════════════════════════════════════

\title{Counted NFT Transfers}

\author{
\IEEEauthorblockN{Qin Wang$^{1,3}$, Minfeng Qi$^{2}$, Guangsheng Yu$^3$, Shiping Chen$^{1}$}\\
\textit{$^1$CSIRO Data61} $|$ \textit{$^2$City University of Macau} $|$ \textit{$^3$University of Technology Sydney} }

\maketitle

% ══════════════════════════════════════════════════════════════════════════════
%  ABSTRACT
% ══════════════════════════════════════════════════════════════════════════════

\begin{abstract}
Non-fungible tokens (NFTs) on Ethereum currently follow a binary mobility paradigm: ERC-721 enables unrestricted transfers, whereas SBTs (ERC-5192) prohibit transfers entirely. We identify a design gap in which no standard mechanism supports \emph{bounded transferability}, where ownership mobility is allowed but limited to a finite number of programmable transfers. We study \textit{counted NFT transfers} and introduce ERC-7634 as a minimal realization compatible with ERC-721. The design augments each token with a transfer counter and configurable cap $L$, allowing ownership to evolve under a finite transfer budget. 
ERC-7634 defines a minimal extension interface with three lightweight functions
(\texttt{transferCountOf}, \texttt{setTransferLimit}, and \texttt{transferLimitOf}),
two events, and native-transfer hooks, requiring fewer than 60 additional lines of Solidity while preserving full backward compatibility with existing NFT infrastructure.

We analyze behavioral and economic consequences of counted transfers. Our results reveal (i) a mobility premium induced by remaining transfer capacity, (ii) a protocol-level \emph{costing signal} that can deter wash trading in cap-aware markets through irreversible budget consumption, (iii) bounded recursive collateralization enabled by limited ownership turnover, and (iv) associated security and gas-cost implications, including wrapper-bypass trade-offs.
Evaluation on calibrated simulations shows that moderate limits (e.g., $L{=}10$) affect fewer than 15\% of tokens under representative transfer distributions, while repeated manipulation becomes unprofitable after a few cycles in a cap-aware pricing model; the additional gas overhead remains below 11\% per transfer.
We further position ERC-7634 within the NFT mobility design space, derive practical cap-selection guidelines, and discuss post-cap ownership outcomes including soulbound conversion, auto-burn, and provenance freeze.
\end{abstract}


% ══════════════════════════════════════════════════════════════════════════════
%  1. INTRODUCTION
% ══════════════════════════════════════════════════════════════════════════════

\section{Introduction}
\label{sec:intro}

The ERC-721 standard~\cite{erc721} established the foundation for \textit{non-fungible tokens} (NFTs)~\cite{wang2021nft} on Ethereum, enabling representation of unique digital assets with unrestricted transferability. Since its adoption, ERC-721 has supported a diverse ecosystem spanning digital art, gaming assets, membership passes, and financial instruments, with cumulative trading volumes exceeding \$68 billion~\cite{nonfungible2024}. Yet the standard assumes that once minted, a token may be transferred unlimited times without constraint or consequence.

This assumption creates a \emph{binary mobility paradigm}. At one extreme, ERC-721 tokens flow freely: any owner can transfer to any recipient at any time. At the other, Soulbound Tokens (SBTs)~\cite{weyl2022decentralized}, formalized through ERC-5192~\cite{erc5192} and ERC-5484~\cite{erc5484}, are permanently non-transferable once bound to an address. Between these poles lies an unaddressed design space: NFTs whose ownership mobility is neither unlimited nor prohibited but bounded by a finite number of ownership transitions. We call this \emph{bounded transferability}, where ownership evolves under a programmable transfer budget.

The absence of bounded transferability creates practical limitations. We show several examples:

\begin{figure}[t]
    \centering
    \includegraphics[width=\linewidth]{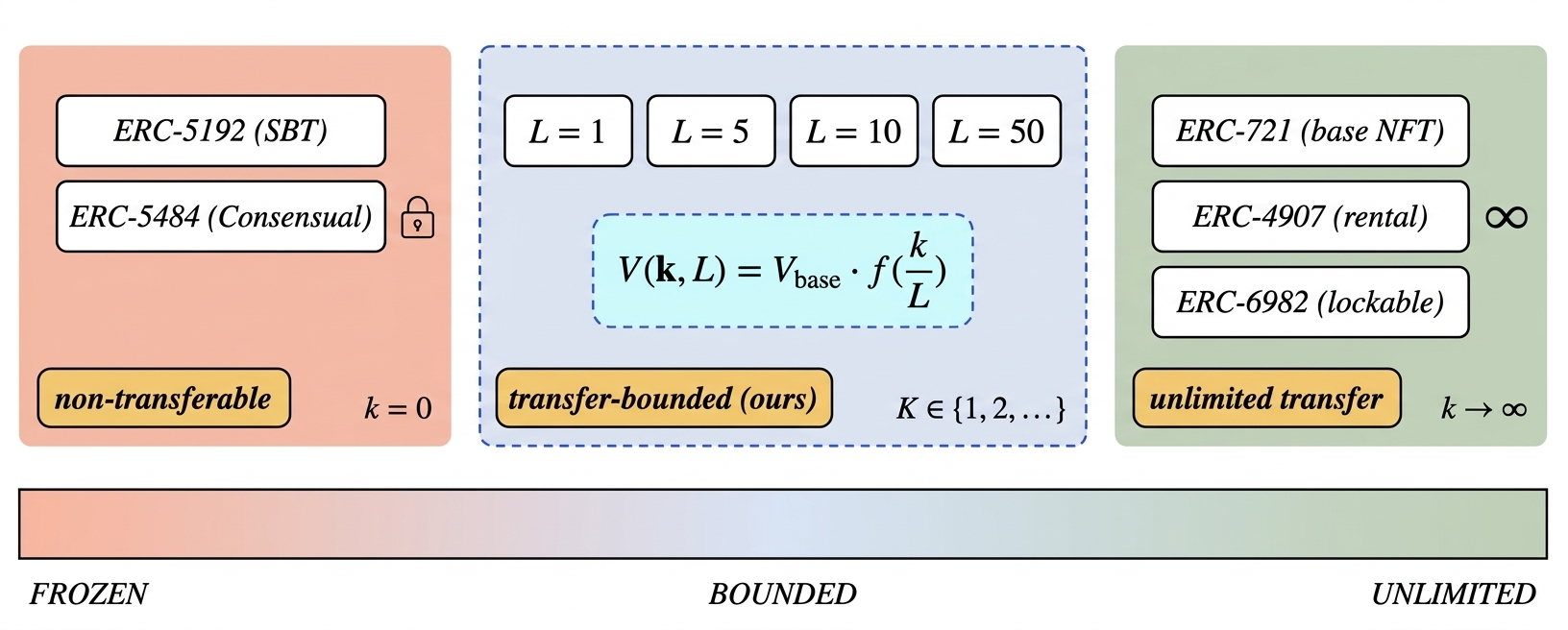}
    \caption{Token mobility. ERC-7634 fills the design gap between non-transferable Soulbound Tokens (left) and fully transferable ERC-721 (right), enabling configurable transfer bounds for gaming, ticketing, memberships, and DeFi collateral.}
    \label{fig:mobility_spectrum}
\end{figure}

\begin{itemize}
\item \textit{Ownership dilution through excessive trading.}
A ``limited edition'' NFT collection preserves supply scarcity (fixed token count) but lacks \emph{transfer scarcity}. Each token can be flipped indefinitely, diluting ownership value. Empirical studies show that 24--80\% of NFT trading volume may constitute wash trading~\cite{von2022nft, das2022understanding}, inflating perceived demand while degrading legitimate market signals.

\item \textit{Unbounded recursive leverage.}
An NFT used as collateral generates a loan that can purchase another NFT, which is then re-collateralized, creating re-hypothecation chains analogous to those amplifying the 2008 financial crisis~\cite{singh2010velocity}. Without transfer bounds, maximum leverage depth is limited only by the loan-to-value (LTV) ratio, approaching $1/(1-\text{LTV})$.

\item \textit{No lifecycle control.}
Many real-world assets evolve through use or transfer~\cite{wang2022exploring,yu2024toward}: event tickets lose validity after resale, software licenses allow limited activations, and physical goods wear with handling. Current NFT standards cannot capture this notion of usage or exhaustion on-chain, forcing application-layer workarounds that fragment the ecosystem.

\end{itemize}

\begin{table*}[t]
\centering
\caption{Comparison of NFT mobility mechanisms across Ethereum token standards}
\label{tab:taxonomy}

\renewcommand{\arraystretch}{1.18}
\footnotesize

\resizebox{\linewidth}{!}{%
\begin{tabular}{l|l|l|c|c|c|c|c|c}
\toprule
\multicolumn{1}{c}{\textbf{Standard}} &
\multicolumn{1}{c}{\textbf{Title}} &
\multicolumn{1}{c|}{\textbf{Primary mechanism}} &
\textbf{Category} &
\multicolumn{1}{c}{\textbf{Transferability}} &
\multicolumn{1}{c}{\textbf{Count aware}} &
\multicolumn{1}{c}{\textbf{Value decay}} &
\textbf{DeFi} &
\textbf{Anti-spec.} \\
\midrule

ERC-721 &
Non-Fungible Token Standard &
Unrestricted ownership transfer &
Base NFT &
Unlimited &
No &
None &
Full &
None \\

ERC-4907 &
Rental NFT Standard &
Time-limited user role separation &
Rental &
Unlimited (split) &
No &
Time-based &
Partial &
Low \\

ERC-5192 &
Minimal Soulbound Token &
Transfer reversion (permanent lock) &
Soulbound &
None &
N/A &
N/A &
None &
Complete \\

ERC-5484 &
Consensual Soulbound Token &
Consent-based binding control &
Consensual SBT &
None (post-bind) &
N/A &
N/A &
None &
Complete \\

ERC-5679 &
Mint and Burn Extension &
Explicit burn lifecycle transition &
Ext. (burn) &
Unlimited + burn &
No &
Binary (burn) &
Partial &
Low \\

ERC-6982 &
Lockable NFT Standard &
Transfer gating via lock state &
Lockable &
Conditional &
No &
None &
Partial &
Medium \\

\midrule
\rowcolor{capWeak}
\textbf{ERC-7634} &
\textbf{Counted Transfer Extension} &
\textbf{Transfer-count enforcement} &
\textbf{Transfer-bounded} &
\textbf{Bounded [0,$L$]} &
\textbf{Yes} &
\textbf{Gradual} &
\textbf{Full} &
\textbf{Med--High} \\
\bottomrule
\end{tabular}%
}
\end{table*}

We study \textit{counted NFT transfers} and realize this capability through ERC-7634~\cite{erc7634}\footnote{The ERC-7634 proposal underwent the standard five-phase Ethereum Improvement Proposal (EIP) process (PR, Draft, Review, Last Call, and Final). Our case for the standardisation process spanned nearly two years (Feb.~2024--late~2025) with active community discussions (\url{https://ethereum-magicians.org/t/erc-7634-limited-transferable-nft/18861}). }, a minimal deployable extension to ERC-721 equipping each token with a transfer counter and configurable cap $L$. Rather than introducing a new token model, the design preserves native ERC-721 semantics and achieves transfer-count awareness through three lightweight functions (\texttt{transferCountOf}, \texttt{setTransferLimit}, and \texttt{transferLimitOf}), associated events, and \emph{native-transfer} hooks (i.e., enforced on direct ERC-721 transfers), showing that counted transfers integrate into existing NFT infrastructure without breaking backward compatibility.

Our approach combines formal specification, systems realization, and empirical evaluation. We formalize counted transfers as a transfer-aware state machine and present a production-ready implementation requiring fewer than 60 lines of additional Solidity atop OpenZeppelin's ERC-721, showing bounded ownership can be enforced with minimal protocol overhead. We then analyze behavioral and economic implications of transfer-bounded ownership and validate these effects through numerical evaluation on 50{,}000 synthetic tokens calibrated to empirical transfer distributions. Finally, we position ERC-7634 within the broader NFT mobility design space and discuss practical post-cap ownership outcomes.

\smallskip
\noindent\textbf{Contributions.}
This paper makes four main contributions.

We identify a fundamental gap in NFT ownership between unrestricted transferability and permanent non-transferability, and formalize \emph{counted transferability} as a transfer-bounded ownership model (\S\ref{sec:design}). We characterize ownership evolution under finite transfer budgets and derive a transfer-aware state abstraction capturing bounded ownership transitions.

We demonstrate that this abstraction can be realized through ERC-7634 (standardized via the Ethereum EIP process), a minimal backward-compatible extension to ERC-721. We present a deployable reference implementation based on OpenZeppelin v5's \texttt{\_update} hook pattern, showing transfer-bounded ownership can be enforced with minimal protocol overhead while preserving existing NFT semantics (\S\ref{sec:design}).

Third, we analytically characterize behavioral, economic, and security implications of counted transfers, including mobility-aware valuation, protocol-level manipulation deterrence, bounded leverage formation, and gas-security trade-offs, and provide a systematic study of wrapper-bypass vulnerability with mitigation strategies (\S\ref{sec:economics}--\S\ref{sec:security}).

Finally, we place ERC-7634 within the broader NFT mobility landscape and discuss deployment implications, including cap-selection insights and post-cap ownership outcomes such as soulbound conversion, auto-burn, lock-and-release, and provenance freeze (\S\ref{sec:taxonomy}--\S\ref{sec:discussion}).

% ══════════════════════════════════════════════════════════════════════════════
%  2. BACKGROUND AND RELATED WORK
% ══════════════════════════════════════════════════════════════════════════════
\section{Background and Related Work}
\subsection{NFT Standards}

ERC-721~\cite{erc721} defines the canonical interface for non-fungible tokens, providing ownership tracking, approval mechanisms, and safe transfer operations. The standard imposes no constraints on ownership transfers, allowing tokens to circulate freely after minting. This design emphasises liquidity and composability, enabling integration with marketplaces, lending protocols, and other DeFi primitives~\cite{jiang2023decentralized,wang2025understandingbrc}.

Subsequent standards extend NFT functionality along complementary dimensions~\cite{qi2025understanding}. ERC-1155~\cite{erc1155} introduces semi-fungible assets, ERC-4907~\cite{erc4907} separates usage rights from ownership for rental scenarios, and ERC-2981~\cite{erc2981} standardizes royalty signaling. Meanwhile, SBTs~\cite{weyl2022decentralized}, realized through ERC-5192~\cite{erc5192} and ERC-5484~\cite{erc5484}, eliminate transferability by binding tokens permanently to an address. Other proposals such as ERC-6982~\cite{erc6982} and ERC-5679~\cite{erc5679} introduce locking, minting, and burning controls.

While these standards expand lifecycle and access-control capabilities, they operate largely orthogonally to ownership mobility: tokens remain either freely transferable or permanently non-transferable, with limited mechanisms governing intermediate ownership transitions.

\subsection{NFT Mobility}

NFT mobility refers to a token's ability to change ownership over time~\cite{wang2021nft}. It plays a central role in market dynamics, shaping how assets circulate, accumulate value, and interact with decentralized applications. Existing standards implicitly define mobility only at two extremes: unrestricted circulation under ERC-721 and permanent immobility in soulbound designs. Yet ownership transitions strongly influence market behavior, protocol leverage, and lifecycle semantics, as repeated transfers enable speculative circulation, recursive collateralization, and prolonged asset reuse.

Despite its importance, mobility has been treated as a binary attribute rather than a controllable protocol dimension. Mechanisms that bound ownership transitions, rather than eliminating or leaving them unconstrained, remain largely unexplored in current NFT standards~\cite{qi2025understanding}.

\subsection{Market Manipulation in NFT Markets}

Wash trading is well documented in NFT markets. It involves trading with oneself to inflate volume metrics. Von Wachter et al.~\cite{von2022nft} report that wash trading constitutes up to 24\% of total NFT trading volume. Das et al.~\cite{das2022understanding} show that wash traders and insiders collude to create pump-and-dump dynamics, with insiders selling during or after wash trading episodes. Huang et al.~\cite{huang2024unveiling} analyze hundreds of millions of NFT transfer transactions, showing that market growth is driven by a relatively small set of dominant participants and exhibits substantial anomalous trading activity. 

% ~\cite{nftdisk2023}

Existing countermeasures rely on post-hoc detection using graph analytics or platform-level policies. Wang et al.~\cite{wang2025manipulation} study manipulation-resilient NFT pricing and demonstrate that strategic trading behaviors can distort valuation signals. ERC-7634 provides a complementary \emph{protocol-level} deterrent by making wash trades consume a finite resource (transfer budget), imposing an economic cost beyond gas fees.

\subsection{Recursive Collateralization}

Recursive collateralization refers to repeated reuse of an asset as loan collateral across multiple borrowing cycles, amplifying systemic leverage and risk~\cite{sugino2025analysis}. In DeFi protocols such as NFTfi, BendDAO, and Blur Lending, NFTs serve as collateral for loans that may be redeemed and re-deposited after ownership transfer, forming recursive leverage loops.

Without ownership transition constraints, leverage depth is primarily bounded by loan-to-value (LTV) ratios, approaching $1/(1-\text{LTV})$ under idealized assumptions. Transfer-count limits in ERC-7634 introduce a structural bound on recursive collateralization by restricting admissible ownership cycles.

\begin{figure}[t]
    \centering
    \includegraphics[width=0.95\linewidth]{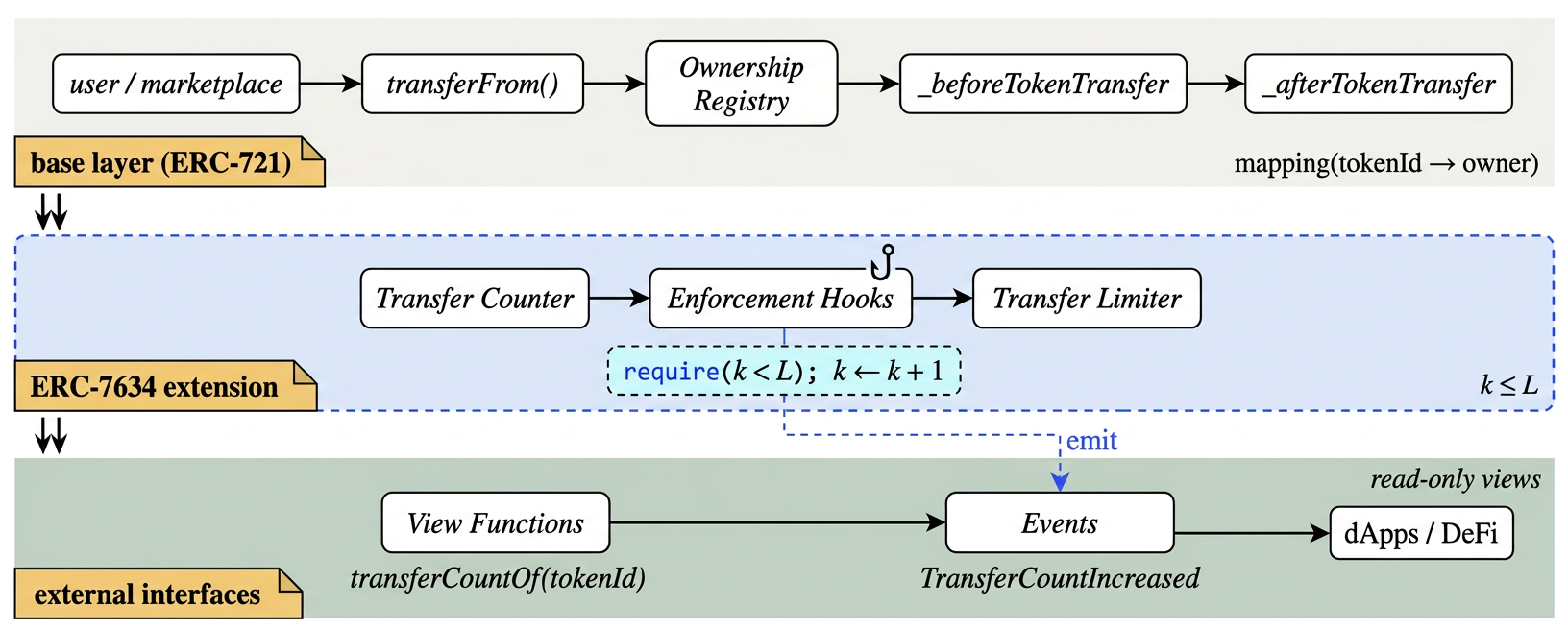}
    \caption{System overview of the ERC-7634 extension. The Transfer Counter, Transfer Limiter, and Enforcement Hooks extend the ERC-721 base layer while exposing read interfaces for external contracts and dApps.}
    \label{fig:system_overview}
\end{figure}

% ══════════════════════════════════════════════════════════════════════════════
%  3. DESIGN AND SPECIFICATION
% ══════════════════════════════════════════════════════════════════════════════

\section{Design and Specification}
\label{sec:design}

\subsection{Design Principles}

ERC-7634 is guided by three principles:

\begin{packeditemize}
    \item \textit{Minimality.} The extension introduces only the interface elements required for transfer-count awareness, avoiding feature bloat and preserving the lightweight nature of ERC-721. Instead of redefining ownership semantics or token structure, ERC-7634 adds a per-token transfer counter and configurable cap enforced through \emph{native-transfer} hooks hooks, ensuring that bounded mobility can be achieved with minimal protocol overhead.

    \item \textit{Backward compatibility.} All ERC-721 semantics are preserved. The standard operates as an opt-in extension layered on top of existing contracts, allowing ERC-721 tokens to function normally unless a transfer limit is explicitly configured. Enforcement occurs only during native transfers (excluding mint and burn), enabling existing marketplaces, wallets, and DeFi protocols to interact with ERC-7634 tokens without modification.

    \item \textit{Composability.} The standard exposes read-only interfaces (\texttt{transferCountOf}, \texttt{transferLimitOf}) and explicit runtime events that allow external protocols to observe transfer-state evolution without altering execution logic. This design enables integration with higher-layer mechanisms such as lending, pricing, lifecycle management, or compliance-aware wrappers, while maintaining ERC-721 interoperability.
\end{packeditemize}

\subsection{Interface Specification}

ERC-7634 defines a lightweight extension interface that introduces transfer-count awareness while preserving ERC-721 semantics. The interface exposes both state-query functions and administrative controls required to enforce bounded transferability at the token level.

\medskip    
\begin{lstlisting}[caption={IERC7634 Interface}, label={lst:interface}]
interface IERC7634 {
  event TransferCountIncreased(
    uint256 indexed tokenId,
    uint256 count
  );
  event TransferLimitUpdated(
    uint256 indexed tokenId,
    uint256 limit
  );
  function transferCountOf(
    uint256 tokenId
  ) external view returns (uint256);
  function setTransferLimit(
    uint256 tokenId,
    uint256 limit
  ) external;
  function transferLimitOf(
    uint256 tokenId
  ) external view returns (uint256);
}
\end{lstlisting}

The interface introduces two observable state variables per token: a \emph{transfer count}, representing the number of successful native ERC-721 ownership transitions, and a \emph{transfer limit}, defining the maximum number of admissible transfers. The function \texttt{transferCountOf} provides a read-only view of the current count, enabling external contracts and analytics systems to track ownership mobility over time. Similarly, \texttt{transferLimitOf} exposes the configured cap, allowing protocols to reason about remaining transfer budget without modifying execution logic.

The function \texttt{setTransferLimit} enables token owners or approved operators to configure or update the transfer bound. Implementations are expected to enforce authorization checks consistent with ERC-721 approval semantics. Importantly, transfer counting applies only to native ownership transfers (i.e., where both sender and recipient are non-zero addresses), ensuring that minting and burning operations do not affect transfer budgets.

Two events provide transparent runtime observability. \texttt{TransferCountIncreased} is emitted after a successful transfer when the counter increments, allowing off-chain indexers and dependent protocols to react to mobility changes. \texttt{TransferLimitUpdated} records administrative updates to transfer constraints, clearly separating configuration changes from runtime ownership transitions.

\subsection{Formal State Machine}

We model each token's transfer state as a tuple $(k, L)$, where $k$ denotes the current transfer count and $L$ the transfer limit. The lifecycle is captured by the following state transitions:

\begin{itemize}
    \item \textit{Mint:} $\bot \rightarrow (0, L_{\text{default}})$, where $L_{\text{default}}=0$ denotes \emph{unbounded transferability} (i.e., no cap), and $L_{\text{default}}>0$ initializes a finite cap.
    
    \item \textit{Limit update:} $(k, L) \rightarrow (k, L')$ with the constraint $L' \geq k$, ensuring the limit cannot be reduced below the accumulated transfer count.
    
    \item \textit{Transfer:} $(k, L) \rightarrow (k+1, L)$ if $(L=0)\ \lor\ (k < L)$, corresponding to a native ERC-721 transfer (both \texttt{from} and \texttt{to} are non-zero). The transition reverts only when $L>0$ and $k \geq L$.
    
    \item \textit{Burn:} $(k, L) \rightarrow \bot$, terminating the token state without modifying the transfer count.
\end{itemize}

The state transition must satisfy the following properties:
\begin{itemize}
   \item \textit{Safety:} For every token $t$ with a finite cap ($L(t)>0$), the transfer counter never exceeds its limit:
$\forall t:\; L(t)>0 \Rightarrow \textsf{transferCount}(t) \leq \textsf{transferLimit}(t)$.

   \item \textit{Liveness:} A native transfer of token $t$ succeeds if and only if $\bigl(L(t)=0 \ \lor\ \textsf{transferCount}(t) < \textsf{transferLimit}(t)\bigr)$
and all ERC-721 transfer preconditions are satisfied.
\end{itemize}

\subsection{Counting Rules}

The transfer count increments \emph{only} on native ERC-721 transfers where both the \texttt{from} and \texttt{to} addresses are non-zero. Minting (\texttt{from} = 0) and burning (\texttt{to} = 0) operations do not affect the count. This ensures that the transfer budget reflects genuine ownership changes rather than lifecycle events.

\subsection{Implementation Pattern}

The reference implementation enforces transfer limits by instrumenting two ERC-721 lifecycle hooks. The logic can be expressed abstractly as the following algorithm.

\begin{algorithm}[H]
\caption{Transfer-Cap Enforcement via ERC-721 Hooks}
\label{alg:hooks}
\small
\textbf{Input:} $(from, to, tokenId)$ \\
\textbf{State:} transferCount[tokenId], transferLimit[tokenId]

\begin{algorithmic}[1]
\If{$from \neq \texttt{0x0}$ \textbf{and} $to \neq \texttt{0x0}$}
    \State \textbf{/* Pre-transfer enforcement */}
    \If{transferLimit[tokenId] $>$ 0 \textbf{and} transferCount[tokenId] $\ge$ transferLimit[tokenId]}
    \State \textbf{revert} ``transfer limit reached''
    \EndIf
\EndIf

\State Execute native ERC-721 transfer

\If{$from \neq \texttt{0x0}$ \textbf{and} $to \neq \texttt{0x0}$}
    \State \textbf{/* Post-transfer accounting */}
    \State transferCount[tokenId] $\leftarrow$ transferCount[tokenId] $+ 1$
    \State emit TransferCountIncreased(tokenId, transferCount[tokenId])
\EndIf
\end{algorithmic}
\end{algorithm}

\subsection{Exemplified Implementation}

Algorithm~\ref{alg:erc7634} presents the core transfer-enforcement logic of a deployable ERC-7634 implementation built atop OpenZeppelin v5's ERC-721. 

The design relies on the \texttt{\_update} internal hook, which replaces the deprecated \texttt{\_beforeTokenTransfer}/ \texttt{\_afterTokenTransfer} pair in OpenZeppelin v5, allowing transfer validation and accounting to be performed atomically within a single execution path. The full contract requires fewer than 60 lines of additional Solidity, showing that ERC-7634 introduces minimal integration overhead for existing ERC-721 deployments.

\begin{algorithm}[t]
\caption{ERC-7634 Transfer Enforcement via \texttt{\_update}}
\label{alg:erc7634}
\small
\textbf{State:}
transferCount[tokenId], transferLimit[tokenId]

\textbf{Input:}
$(to, tokenId, auth)$

\begin{algorithmic}[1]
\State $from \leftarrow ownerOf(tokenId)$

\If{$from \neq 0x0$ \textbf{and} $to \neq 0x0$}
    \Comment{native transfer (exclude mint/burn)}
    \If{transferLimit[tokenId] $>$ 0 \textbf{and} transferCount[tokenId] $\ge$ transferLimit[tokenId]}
    \State \textbf{revert} ``transfer limit reached''
\EndIf
\EndIf

\State $prev \leftarrow$ Execute ERC-721 native update$(to, tokenId, auth)$

\If{$from \neq 0x0$ \textbf{and} $to \neq 0x0$}
    \State $transferCount[tokenId] \leftarrow transferCount[tokenId] + 1$
    \State emit TransferCountIncreased$(tokenId,$
           $transferCount[tokenId])$
\EndIf

\State \Return $prev$
\end{algorithmic}
\end{algorithm}

The key design choice is consolidating both the guard check and the counter increment inside \texttt{\_update}, which is invoked for every mint, transfer, and burn operation. As shown in Algorithm~\ref{alg:erc7634}, the condition \texttt{from != address(0) \&\& to != address(0)} isolates native ownership transfers, ensuring that only genuine transfers consume the transfer budget while mint and burn operations remain count-neutral.

% ══════════════════════════════════════════════════════════════════════════════
%  4. ECONOMIC ANALYSIS
% ══════════════════════════════════════════════════════════════════════════════

\section{Economic Analysis}
\label{sec:economics}

\subsection{Transfer-Adjusted Valuation Model}

Transfer caps introduce a new economic dimension: \emph{mobility}, defined as the remaining capacity of a token to change ownership. We model token value as a function of remaining transfer capacity. Unless stated otherwise, the analysis below focuses on the capped case $L>0$; the uncapped case $L=0$ reduces to $V=V_{\text{base}}$.

\begin{definition}[Mobility Premium]
Let $V_{\text{base}}$ denote the intrinsic value of a token with unlimited transferability.  For a token with transfer limit $L$ and current count $k$, define remaining transfers $r=L-k$ when $L>0$. The transfer-adjusted value is
\begin{equation}
V(k,L)=
\begin{cases}
V_{\text{base}}, & L=0 \ (\text{unbounded}),\\[2pt]
V_{\text{base}}\cdot f\!\left(\frac{L-k}{L}\right), & L>0,
\end{cases}
\label{eq:valuation}
\end{equation}
where $f:[0,1]\rightarrow[\rho,1]$ is a \emph{mobility premium function} satisfying $f(1)=1$, and $\rho\in[0,1)$ captures an optional residual terminal value when mobility is exhausted (i.e., $f(0)=\rho$). Unless otherwise stated, we use $\rho=0$.

\end{definition}

We in this paper consider four representative classes of premium functions:

\begin{enumerate}
    \item \textit{Linear:} $f(x)=x$, where value decays proportionally with each transfer.
    
    \item \textit{Concave ($\gamma<1$):} $f(x)=x^\gamma$, where early transfers incur small value loss while later transfers become increasingly costly, capturing option-like behavior in which final ownership transitions are most valuable.
    
    \item \textit{Convex ($\gamma>1$):} $f(x)=x^\gamma$, where value declines rapidly after initial transfers and then stabilizes, modeling assets with strong first-owner or ``freshness'' premiums.
    
    \item \textit{Threshold ($\tau$):} $f(x)=\max\!\bigl(0,(x-\tau)/(1-\tau)\bigr)$ for $x>\tau$, and a residual value otherwise, representing assets whose value collapses once remaining mobility falls below a critical fraction.
\end{enumerate}

\subsection{Marginal Mobility Cost}

Transfer-bounded ownership introduces a per-transfer economic friction. We define \emph{marginal mobility cost} as the loss incurred by consuming one additional transfer opportunity.

\begin{equation}
\text{MC}(k,L)=V(L-k,L)-V(L-k-1,L),
\label{eq:marginal_cost}
\end{equation}
which measures the decrease in token value when the remaining transfer capacity decreases from $r=L-k$ to $r-1$.

Under the concave premium model $f(x)=x^\gamma$ with $\gamma=0.5$, the marginal mobility cost becomes
\begin{equation}
\text{MC}_{\text{concave}}(k,L)
=V_{\text{base}}
\left[
\sqrt{\frac{L-k}{L}}
-\sqrt{\frac{L-k-1}{L}}
\right].
\end{equation}

Because $f$ is concave, $\text{MC}(k,L)$ increases as the remaining transfer capacity decreases, implying that later transfers destroy disproportionately more value than earlier ones. Economically, this induces an endogenous friction on excessive trading: rational agents internalize the increasing opportunity cost of mobility consumption and therefore reserve transfers for higher-value ownership changes.

Table~\ref{tab:marginal_cost} in Section~\ref{sec:evaluation} empirically quantifies this effect and illustrates how mobility scarcity translates into progressively increasing transfer costs.

\subsection{Wash Trading Deterrence}

We formalize how transfer caps reshape the incentive structure of wash trading by introducing an irreversible economic cost tied to ownership mobility.

\begin{proposition}[Wash Trading Deterrence in a Cap-Aware Pricing Model]
\label{thm:wash}
Under the concave valuation model with $\gamma=0.5$, consider a \emph{cap-aware market} in which buyers observe remaining transfer counts on-chain and price tokens accordingly. The attacker's profit after performing $n$ wash trades is
\begin{equation}
\Pi_{\text{cap}}(n,L)
= V(L-n,L)\cdot(1+\alpha) - V_{\text{base}} - n g,
\end{equation}
where $\alpha$ denotes the artificial price inflation induced by wash trading, $g$ is the per-trade transaction cost, and
\begin{equation}
V(L-n,L)=V_{\text{base}}\cdot\sqrt{\frac{L-n}{L}}
\end{equation}
is the transfer-adjusted fair value after $n$ transfers are consumed. Wash trading becomes unprofitable ($\Pi_{\text{cap}}\le 0$) when
\begin{equation}
n \ge n^* \;\text{s.t.}\;
V_{\text{base}}\sqrt{\frac{L-n^*}{L}}(1+\alpha)
\le V_{\text{base}} + n^* g .
\end{equation}
\end{proposition}

The key design is that each wash trade permanently consumes mobility by reducing the remaining transfer budget. Unlike traditional markets, where manipulative trades incur temporary gas or fee costs, ERC-7634 introduces a state-dependent cost that irreversibly decreases fundamental value. Rational buyers internalize this loss by discounting tokens with lower remaining transfer capacity, converting wash trading from a repeatable manipulation into a self-limiting process.

For comparison, without transfer caps, the attacker's profit reduces to
\begin{equation}
\Pi_{\text{nocap}} = V_{\text{base}}\cdot\alpha - n g,
\end{equation}
which remains positive as long as artificial price inflation exceeds transaction costs. For example, with $\alpha=0.3$ and $V_{\text{base}}=10$ ETH, the attacker gains approximately $3.0$ ETH absent mobility constraints. Introducing a transfer cap fundamentally alters this outcome: when $L=10$, wash trading becomes unprofitable after roughly five trades, while for $L=5$ profitability disappears after only three trades. This shows that, under cap-aware pricing, bounded transferability can act as a protocol-level deterrent by turning repeated manipulation into irreversible value depreciation.

\begin{figure}[t]
    \centering
    \includegraphics[width=\linewidth]{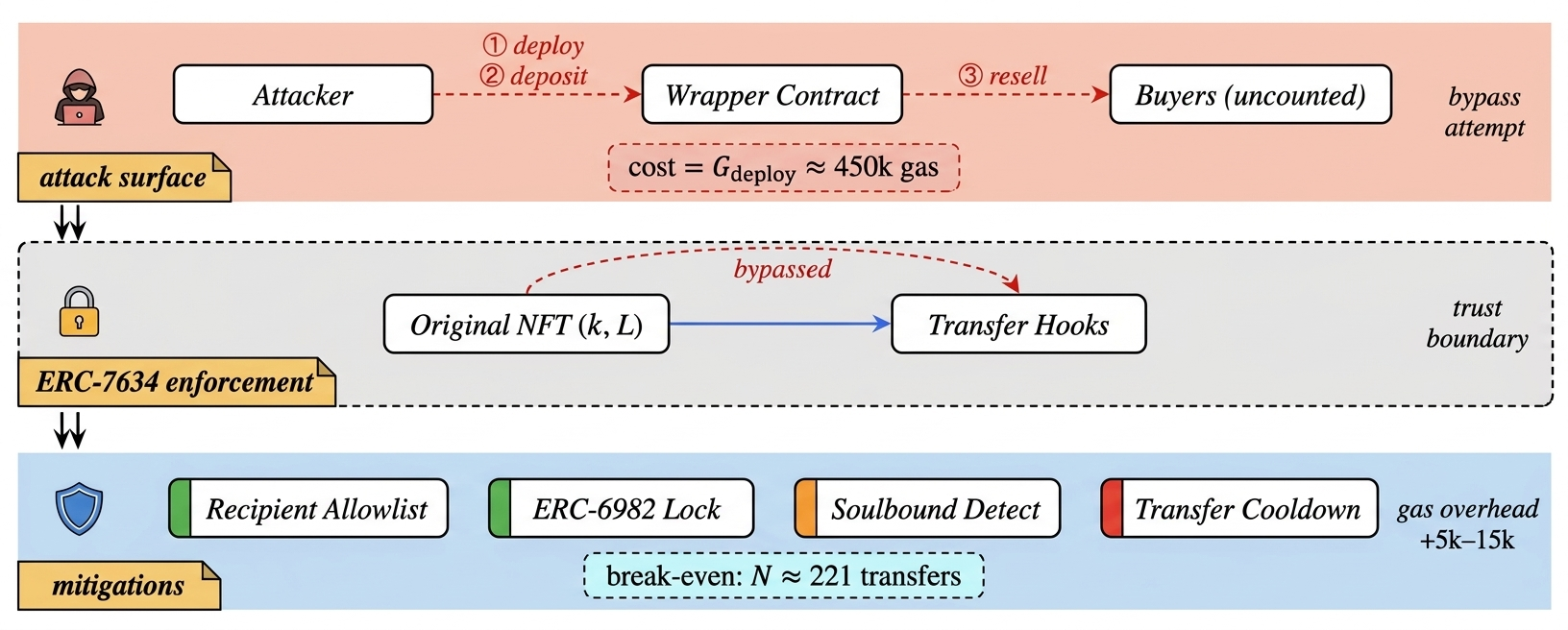}
    \caption{Wrapper bypass threat model. An attacker deploys a wrapper contract to hold the original NFT (consuming one transfer), then transfers wrapper ownership freely, bypassing ERC-7634's transfer cap. Mitigations are on the right.}
    \label{fig:threat_model}
\end{figure}

\subsection{Recursive Leverage Bounding}

In DeFi lending protocols, recursive leverage arises when an NFT is repeatedly re-used as collateral across borrowing cycles. Under a conservative custody-by-transfer model, each re-hypothecation cycle requires at least two native transfers: one to deposit the NFT into a lending contract and another to redeem, liquidate, or reposition the asset. Under a transfer cap $L$, the maximum attainable re-hypothecation depth is therefore bounded by
\begin{equation}
d_{\max}=\left\lfloor\frac{L}{2}\right\rfloor .
\end{equation}

Given a loan-to-value ratio $\text{LTV}$, recursive borrowing forms a geometric leverage process. The maximum leverage becomes
\begin{equation}
\Lambda_{\max}(L)=\sum_{i=0}^{d_{\max}}\text{LTV}^i
=\frac{1-\text{LTV}^{\,d_{\max}+1}}{1-\text{LTV}}.
\end{equation}

Unlike conventional NFT markets, where leverage depth is theoretically unbounded and converges to $1/(1-\text{LTV})$, transfer limits introduce a protocol-level structural constraint that truncates the geometric series. Consequently, leverage amplification becomes a function of mobility scarcity rather than purely financial parameters.

For example, with $\text{LTV}=0.7$ and $L=10$, we obtain $d_{\max}=5$, yielding $\Lambda_{\max}\approx2.94\times$, compared to the unbounded theoretical maximum $1/(1-0.7)=3.33\times$. This corresponds to an 11.7\% reduction in achievable leverage. Smaller transfer budgets impose stronger bounds: when $L=6$, leverage decreases to $\Lambda_{\max}\approx2.53\times$, representing a 24.0\% reduction.

% ══════════════════════════════════════════════════════════════════════════════
%  5. SECURITY ANALYSIS
% ══════════════════════════════════════════════════════════════════════════════

\section{Security Analysis}
\label{sec:security}

\subsection{Wrapper Bypass Attack}

The primary security limitation of ERC-7634 arises from the \emph{wrapper bypass}. Because transfer limits apply only to native ERC-721 transfers of the underlying token, ownership can be indirectly reassigned through an intermediary contract without consuming additional transfer budget, as illustrated in Fig.\ref{fig:threat_model}.

\begin{definition}[Wrapper Bypass]
Let token $t$ have transfer limit $L$. An attacker deploys a wrapper contract $W$ and transfers $t$ into $W$, consuming one native transfer. Control of $t$ is then reassigned by transferring ownership of $W$ (e.g., via a wrapper NFT or administrative key), which is not accounted for by ERC-7634's transfer counter. The effective bypass cost is
\begin{equation}
C_{\text{bypass}}
= G_{\text{deploy}} + G_{\text{deposit}} + n \cdot G_{\text{wrapper\_transfer}},
\end{equation}
where $G_{\text{deploy}} \approx 450{,}000$ gas denotes the deployment cost of the wrapper contract.
\end{definition}

Conceptually, the attack exploits the distinction between \emph{asset transfer} and \emph{control transfer}. While ERC-7634 constrains ownership mobility at the token layer, composability in Ethereum allows control rights to be reassigned at higher abstraction layers. As a result, wrapper ownership transfers can simulate secondary trading without triggering transfer-count enforcement (Fig.\ref{fig:threat_model}).

\subsection{Budget vs.\ Enforcement Model}

The wrapper bypass highlights a fundamental design distinction. ERC-7634 provides a \textbf{budgeting primitive} that exposes remaining mobility (``this token has $r$ transfers left'') rather than an enforcement guarantee ensuring the token cannot exceed $L$ transfers under all execution paths.

This distinction reflects a design trade-off between strict enforcement and ecosystem composability. Absolute prevention would require restricting contract interactions or wrapper constructions, which would conflict with ERC-721 interoperability and the permissionless composability of Ethereum. Instead, ERC-7634 introduces an economically meaningful signal that markets and protocols can interpret and price.

We argue that a budgeting model is appropriate for most practical deployments:

\begin{packeditemize}
    \item \textit{Voluntary compliance ecosystems} (gaming, memberships, digital art). Participants value transparent mobility signals, and wrapper bypass becomes socially visible and reputationally costly, reducing its practical attractiveness.
    
    \item \textit{Economic deterrence}. Wrapper deployment incurs a non-trivial fixed cost (approximately 450k gas, $\approx\$40$ at 30 gwei), creating a baseline barrier that discourages manipulation or excessive trading for low- and mid-value assets.
    
    \item \textit{Protocol-level enforcement}. Applications requiring strict guarantees (e.g., regulated assets, collateral DeFi protocols) can build additional enforcement layers such as compliant wrappers, registries, or allowlisted transfer paths, extending ERC-7634 without breaking its minimal core design.
\end{packeditemize}
    
In this sense, ERC-7634 should be interpreted as a \emph{coordination primitive}: it standardizes observable mobility constraints while allowing higher-layer systems to choose stronger enforcement mechanisms according to their security and governance requirements~\cite{wang2025understanding}.

\subsection{Mitigation Strategies}

We evaluate five representative mitigation strategies against wrapper bypass, each reflecting a different trade-off between enforcement strength, composability, and execution overhead.

\begin{packeditemize}
    \item \textit{Recipient allowlist.}
    Transfers are restricted to pre-approved addresses, preventing deposits into arbitrary wrapper contracts. This approach provides strong bypass resistance but significantly reduces DeFi composability and permissionless interaction. Additional gas cost: $\sim$8{,}200 per transfer.

    \item \textit{Soulbound wrapper detection.}
    The contract verifies whether the recipient is a contract address and attempts to detect re-transfer functionality. This provides moderate resistance; however, adversaries can design wrappers that evade heuristic detection. Additional gas cost: $\sim$12{,}400.

    \item \textit{ERC-6982 lockable integration.}
    The NFT is locked within the receiving contract, preventing extraction without consuming a counted transfer. This achieves strong resistance while preserving partial composability, at the cost of additional protocol complexity. Additional gas cost: $\sim$15{,}600.

    \item \textit{Transfer cooldown period.}
    A minimum time interval is enforced between successive transfers. While this does not prevent wrapping, it reduces high-frequency flipping and automated manipulation. Additional gas cost: $\sim$5{,}100.

    % \item \textit{No mitigation (baseline).}
    % The system adopts the budgeting model without additional enforcement, relying on market transparency and economic incentives. This option preserves full composability and is suitable for open ecosystems where mobility signaling alone provides sufficient deterrence.
\end{packeditemize}

% ══════════════════════════════════════════════════════════════════════════════
%  6. NUMERICAL EVALUATION
% ══════════════════════════════════════════════════════════════════════════════

\section{Evaluation}
\label{sec:evaluation}

We validate the analytical models through numerical simulations based on synthetic datasets calibrated to published empirical statistics on Ethereum NFT transfer patterns.

\subsection{Experimental Setup}

We simulate 50{,}000 NFTs spanning five representative collection categories: profile-picture (PFP) collections (e.g., Bored Ape Yacht Club), generative art (e.g., Art Blocks), gaming assets, membership passes, and metaverse land. Transfer frequencies follow power-law distributions with collection-specific parameters (Table~\ref{tab:transfer_stats}), aligned with prior empirical measurements~\cite{von2022nft, das2022understanding}. Unless otherwise specified, simulations assume a baseline token value $V_{\text{base}} = 10$ ETH and a per-transfer transaction cost of $g = 0.005$ ETH.

\textit{Transfer-budget parameter.}
ERC-7634 introduces a transfer budget $L$, representing the maximum
number of ownership transitions permitted during an NFT’s lifetime.
Unless otherwise specified, we evaluate three representative values
$L \in \{5,10,15,20,50\}$.
These values are not arbitrary tuning parameters, but correspond to
\emph{strong}, \emph{moderate}, and \emph{weak} transfer-bounded
ownership settings calibrated to empirically observed NFT transfer
distributions, where the majority of tokens experience only a small
number of lifetime transfers.

\subsection{Transfer Distribution Analysis}

% \begin{table}[t]
% \centering
% \caption{Transfer Count Distribution Statistics by Collection Type (10{,}000 tokens per collection)}
% \label{tab:transfer_stats}
% \begin{tabular}{c|cc|ccccc}
% \toprule
% \textbf{Collection} & \textbf{Mean} & \textbf{Med.} & \textbf{P90} & \textbf{P95} & \textbf{P99} \\
% \midrule
% PFP (BAYC-like)     & 6.30  & 1 & 9   & 19  & 83  \\
% Art (ArtBlocks)     & 2.62  & 1 & 4   & 7   & 23  \\
% Gaming Items        & 12.83 & 2 & 17  & 41  & 304 \\
% Memberships         & 1.60  & 1 & 3   & 4   & 9   \\
% Metaverse Land      & 4.10  & 1 & 6   & 11  & 50  \\
% \bottomrule
% \end{tabular}
% \end{table}

\begin{table}[t]
\centering
\small
\renewcommand{\arraystretch}{1.18}
\setlength{\tabcolsep}{6pt}

\caption{Transfer count distribution statistics by collection type
(computed over 10{,}000 sampled tokens per collection)}
\label{tab:transfer_stats}

\begin{tabular}{c|cc| r r r}
\toprule
\textbf{Collection}
& \textbf{Mean}
& \textbf{Median}
& \cellcolor{capWeak}\textbf{P90}
& \cellcolor{capMedium}\textbf{P95}
& \cellcolor{capStrong}\textbf{P99} \\
\midrule

\rowcolor{rowlabel}
PFP (BAYC-like)
& 6.30  & 1 & 9   & 19  & 83  \\

Art (ArtBlocks)
& 2.62  & 1 & 4   & 7   & 23  \\

\rowcolor{rowlabel}
Gaming Items
& \textbf{12.83} & 2 & \textbf{17} & \textbf{41} & \textbf{304} \\

Memberships
& 1.60  & 1 & 3   & 4   & 9   \\

\rowcolor{rowlabel}
Metaverse Land
& 4.10  & 1 & 6   & 11  & 50  \\

\bottomrule
\end{tabular}
\end{table}

Table~\ref{tab:transfer_stats} shows that transfer activity is highly right-skewed across all categories. Median transfer counts remain close to one, whereas extreme tail values reach hundreds of transfers for gaming assets. This heavy-tailed structure implies that transfer caps primarily affect a small subset of high-frequency tokens rather than typical ownership behavior.

% \begin{table}[t]
% \centering
% \caption{Percentage of Tokens Exceeding Transfer Cap by Collection Type}
% \label{tab:cap_impact}
% \begin{tabular}{c|cccccc}
% \toprule
% \textbf{Collection} & \textbf{3} & \textbf{5} & \textbf{10} & \textbf{20} & \textbf{50} & \textbf{100} \\
% \midrule
% PFP          & 24.1 & 16.0 & 8.8  & 4.4  & 1.7 & 0.9 \\
% Art          & 13.1 & 7.0  & 2.9  & 1.2  & 0.3 & 0.1 \\
% Gaming       & 32.4 & 23.3 & 14.2 & 8.5  & 4.3 & 2.5 \\
% Memberships  & 6.0  & 2.6  & 0.8  & 0.2  & 0.0 & 0.0 \\
% Metaverse    & 18.8 & 11.7 & 5.5  & 2.5  & 1.0 & 0.4 \\
% \bottomrule
% \end{tabular}
% \end{table}

\begin{table}[t]
\centering
\small
\renewcommand{\arraystretch}{1.15}
\setlength{\tabcolsep}{7pt}

\caption{Tokens exceeding transfer caps by collection type.}
\label{tab:cap_impact}

\begin{tabular}{c| cc cc cc}
\toprule
\multirow{2}{*}{\textbf{Collection}} &
\multicolumn{6}{c}{\textbf{Transfer Cap ($L$)}} \\
\cmidrule(lr){2-7}
& \cellcolor{capStrong}\textbf{3}
& \cellcolor{capStrong}\textbf{5}
& \cellcolor{capMedium}\textbf{10}
& \cellcolor{capMedium}\textbf{20}
& \cellcolor{capWeak}\textbf{50}
& \cellcolor{capWeak}\textbf{100} \\
\midrule

\rowcolor{rowlabel} PFP         
& 24.1 & 16.0 & 8.8  & 4.4  & 1.7 & 0.9 \\

Art         
& 13.1 & 7.0  & 2.9  & 1.2  & 0.3 & 0.1 \\

\rowcolor{rowlabel} Gaming      
& 32.4 & 23.3 & 14.2 & 8.5  & 4.3 & 2.5 \\

Memberships 
& 6.0  & 2.6  & 0.8  & 0.2  & 0.0 & 0.0 \\

\rowcolor{rowlabel} Metaverse   
& 18.8 & 11.7 & 5.5  & 2.5  & 1.0 & 0.4 \\

\bottomrule
\end{tabular}
\end{table}

Table~\ref{tab:cap_impact} quantifies the impact of different transfer caps across NFT collections. A moderate cap ($L=10$) affects fewer than 15\% of tokens in all categories under our calibrated transfer distributions, including gaming assets (14.2\%), indicating that transfer-bounded designs primarily limit high-frequency transfers while leaving the majority of typical ownership changes unaffected.

\begin{figure}[t]
    \centering
    \begin{subfigure}[b]{0.48\columnwidth}
        \centering
        \includegraphics[width=\textwidth]{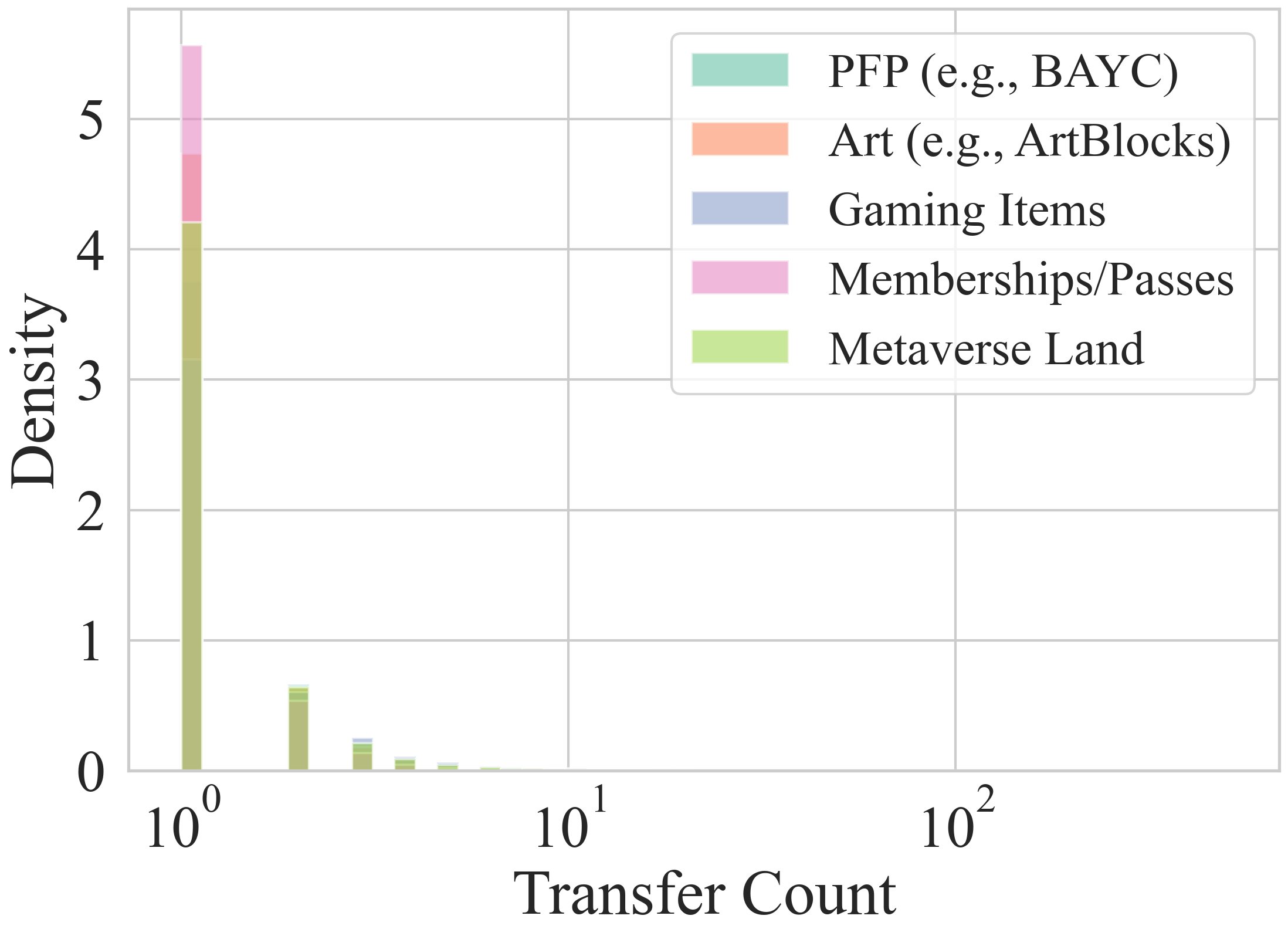}
        \caption{Probability density (/log).}
        \label{fig:transfer_dist_a}
    \end{subfigure}
    \hfill
    \begin{subfigure}[b]{0.48\columnwidth}
        \centering
        \includegraphics[width=\textwidth]{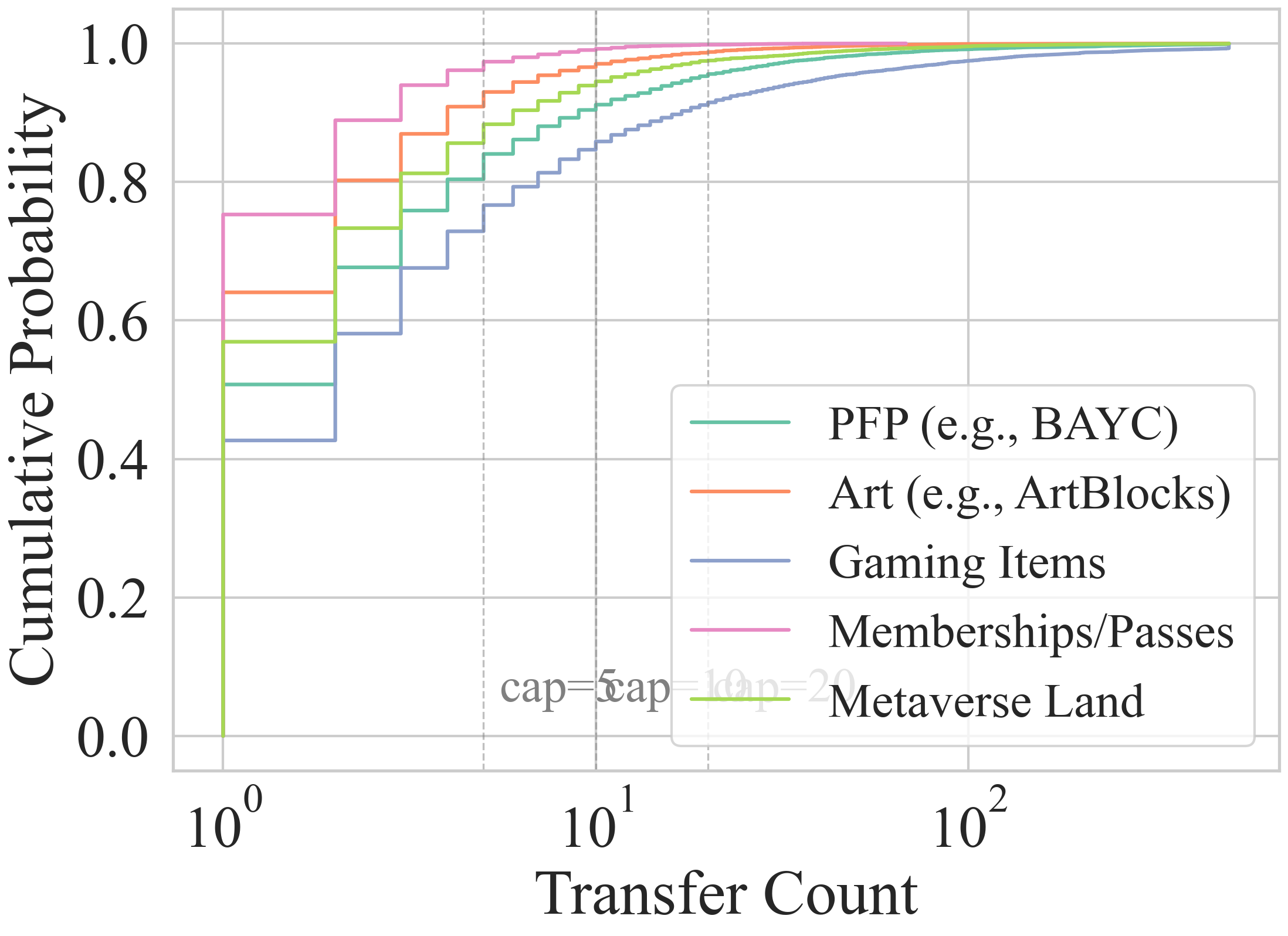}
        \caption{CDF with cap thresholds.}
        \label{fig:transfer_dist_b}
    \end{subfigure}
    \caption{Transfer count distributions across collection types. Cap thresholds at $L \in \{5,10,20\}$ are shown as dashed lines.}
    \label{fig:transfer_dist}
\end{figure}

\begin{figure}[t]
    \centering
    \includegraphics[width=0.9\columnwidth]{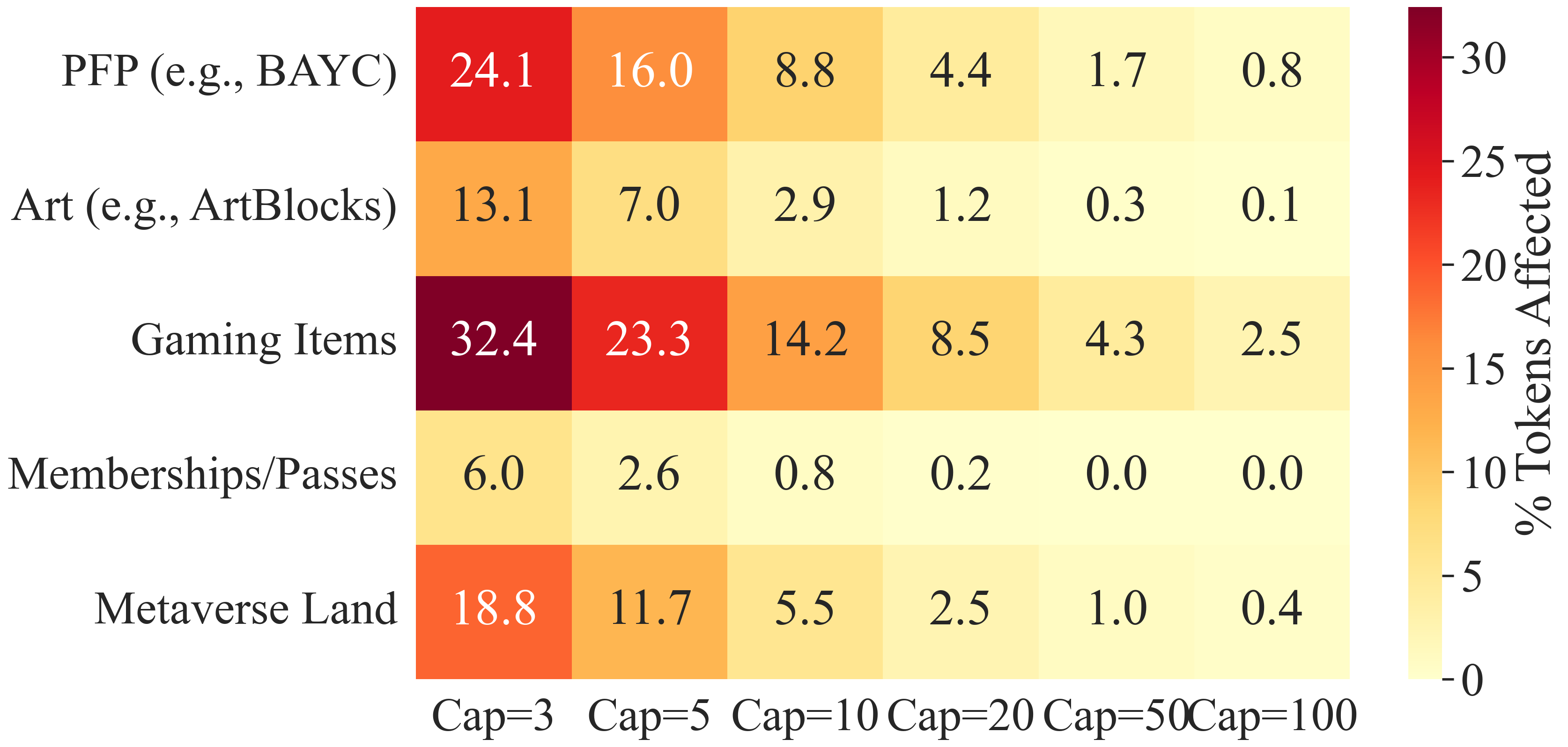}
    \caption{Token percentages exceeding each transfer cap by collection type. Darker cells indicate stronger cap impact.}
    \label{fig:heatmap}
\end{figure}

Fig.\ref{fig:transfer_dist} further illustrates the power-law nature of transfer activity. Estimated exponents vary significantly across collections: gaming items exhibit the heaviest tail ($\alpha \approx 1.8$, P99 = 304 transfers), reflecting active in-game trading, whereas memberships show the steepest decay ($\alpha \approx 3.0$, P99 = 9), consistent with assets rarely resold. These differences directly inform cap design. A uniform cap of $L=10$ constrains only 0.8\% of membership tokens but 14.2\% of gaming assets, suggesting that collection-specific parameterization is necessary for balanced policy outcomes. Fig.\ref{fig:heatmap} summarizes this cap-sensitivity landscape.

\smallskip
\begin{center}
\fbox{%
\begin{minipage}{0.9\linewidth}
\textbf{Takeaway 1: Transfer caps target the tail.}
Transfer caps primarily constrain the speculative long tail of ownership activity. A cap of $L{=}10$ leaves more than 85\% of tokens unaffected across all collection types while bounding the heavy-tailed dynamics associated with wash trading and excessive turnover.
\end{minipage}}
\vspace{5pt}
\end{center}

\subsection{Valuation Model Results}

% \begin{table}[t]
% \centering
% \caption{Token Valuation (ETH) Under Four Mobility Premium Models ($V_{\text{base}} = 10$ ETH, $L = 20$).}
% \label{tab:valuation}
% \begin{tabular}{cc|ccc|c}
% \toprule
% \textbf{$k$} & \textbf{$k/L$} & \textbf{Linear} & \textbf{Concave} & \textbf{Convex} & \textbf{Threshold} \\
% \midrule
% 20 & 1.00 & 10.00 & 10.00 & 10.00 & 10.00 \\
% 18 & 0.90 & 9.00  & 9.49  & 8.10  & 8.75  \\
% 15 & 0.75 & 7.50  & 8.66  & 5.63  & 6.88  \\
% 10 & 0.50 & 5.00  & 7.07  & 2.50  & 3.75  \\
% 5  & 0.25 & 2.50  & 5.00  & 0.63  & 0.63  \\
% 2  & 0.10 & 1.00  & 3.16  & 0.10  & 0.50  \\
% 0  & 0.00 & 0.00  & 0.00  & 0.00  & 0.50  \\
% \bottomrule
% \end{tabular}
% \end{table}

\begin{table}[t]
\centering
\small
\renewcommand{\arraystretch}{1.18}
\setlength{\tabcolsep}{6pt}

\caption{{Token Valuation Under Mobility Premium Models}
($V_{\text{base}} = 10$ ETH, $L = 20$). }
\label{tab:valuation}

\begin{tabular}{c c | c|cc|c}
\toprule
\textbf{$k$} & \multicolumn{1}{c|}{\textbf{$k/L$}}
& \multicolumn{1}{c|}{\textbf{Linear}}
&\multicolumn{1}{c}{ \textbf{Concave}}
&\multicolumn{1}{c|}{ \textbf{Convex}}
& \textbf{Threshold} \\
\midrule

\rowcolor{capWeak}
20 & 1.00 & 10.00 & 10.00 & 10.00 & 10.00 \\

18 & 0.90 & 9.00  & 9.49  & 8.10  & 8.75  \\

\rowcolor{rowlabel}
15 & 0.75 & 7.50  & 8.66  & 5.63  & 6.88  \\

10 & 0.50 & 5.00  & 7.07  & 2.50  & 3.75  \\

\rowcolor{capMedium}
5  & 0.25 & 2.50  & 5.00  & 0.63  & 0.63  \\

2  & 0.10 & 1.00  & 3.16  & 0.10  & 0.50  \\

\rowcolor{capStrong}
0  & 0.00 & 0.00  & 0.00  & 0.00  & 0.50  \\

\bottomrule
\end{tabular}
\end{table}

\begin{table}[t]
\centering
\small
\renewcommand{\arraystretch}{1.18}
\setlength{\tabcolsep}{7pt}

\caption{{Marginal Mobility Cost as \% of $V_{\text{base}}$} (Concave Model, Selected Values).}
\label{tab:marginal_cost}

\begin{tabular}{ >{\columncolor{capWeak}}c|r r r r}
\toprule
\textbf{Transfer stage} 
& \textbf{$L=5$}
& \textbf{$L=10$}
& \textbf{$L=20$}
& \textbf{$L=50$} \\
\midrule

% \rowcolor{capWeak}
First transfer 
& 10.6\% & 5.1\%  & 2.5\%  & 1.0\% \\

% \rowcolor{capMedium}
Mid-point 
& 14.2\% & 10.1\% & 7.1\%  & 4.5\% \\

% \rowcolor{capStrong}
Last transfer 
& 44.7\% & 31.6\% & 22.4\% & 14.1\% \\

\bottomrule
\end{tabular}
\end{table}

\begin{figure}[t]
    \centering
    \begin{subfigure}[b]{0.48\columnwidth}
        \centering
        \includegraphics[width=\textwidth]{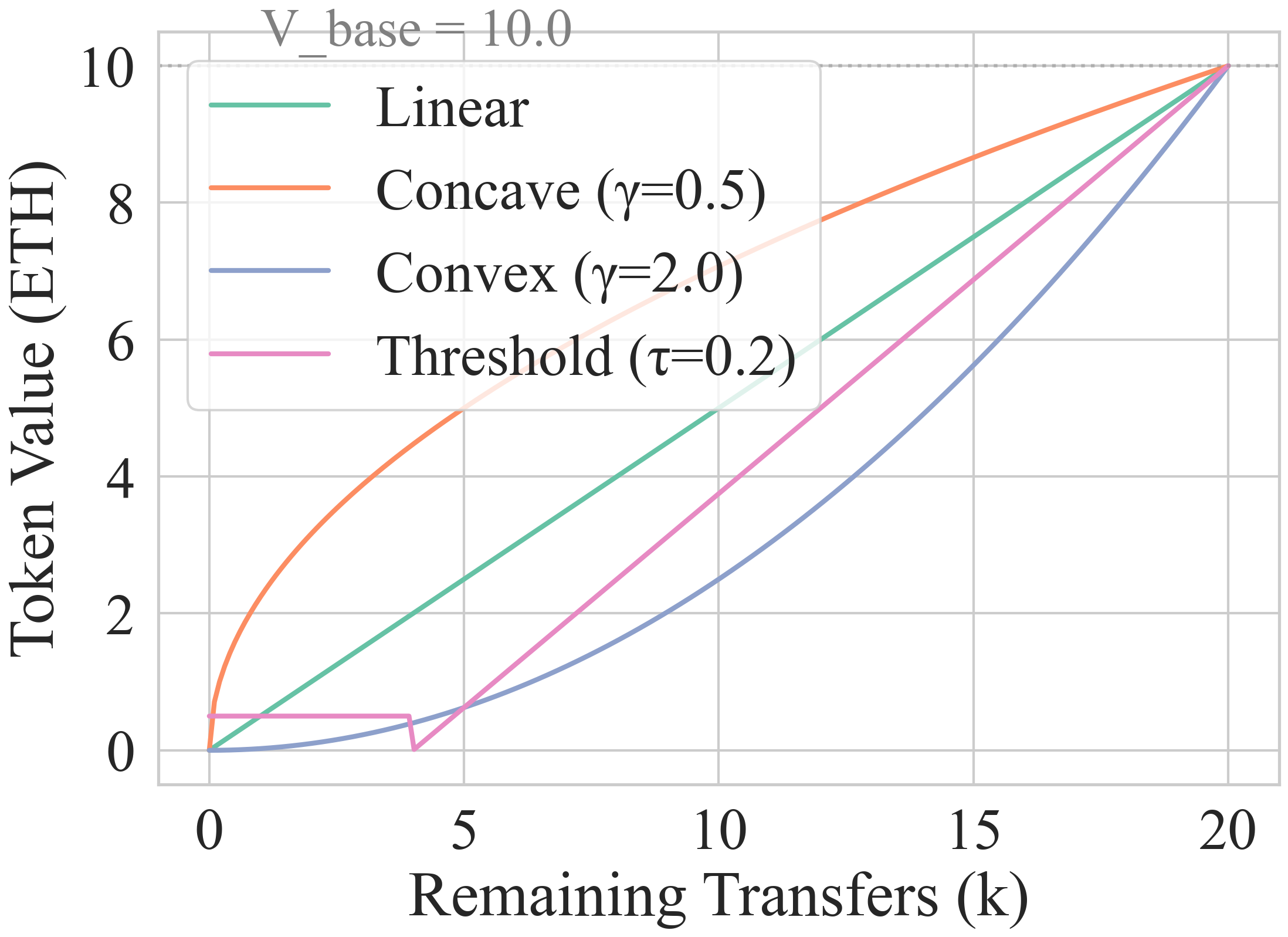}
        \caption{Four valuation models.}
        \label{fig:valuation_a}
    \end{subfigure}
    \hfill
    \begin{subfigure}[b]{0.48\columnwidth}
        \centering
        \includegraphics[width=\textwidth]{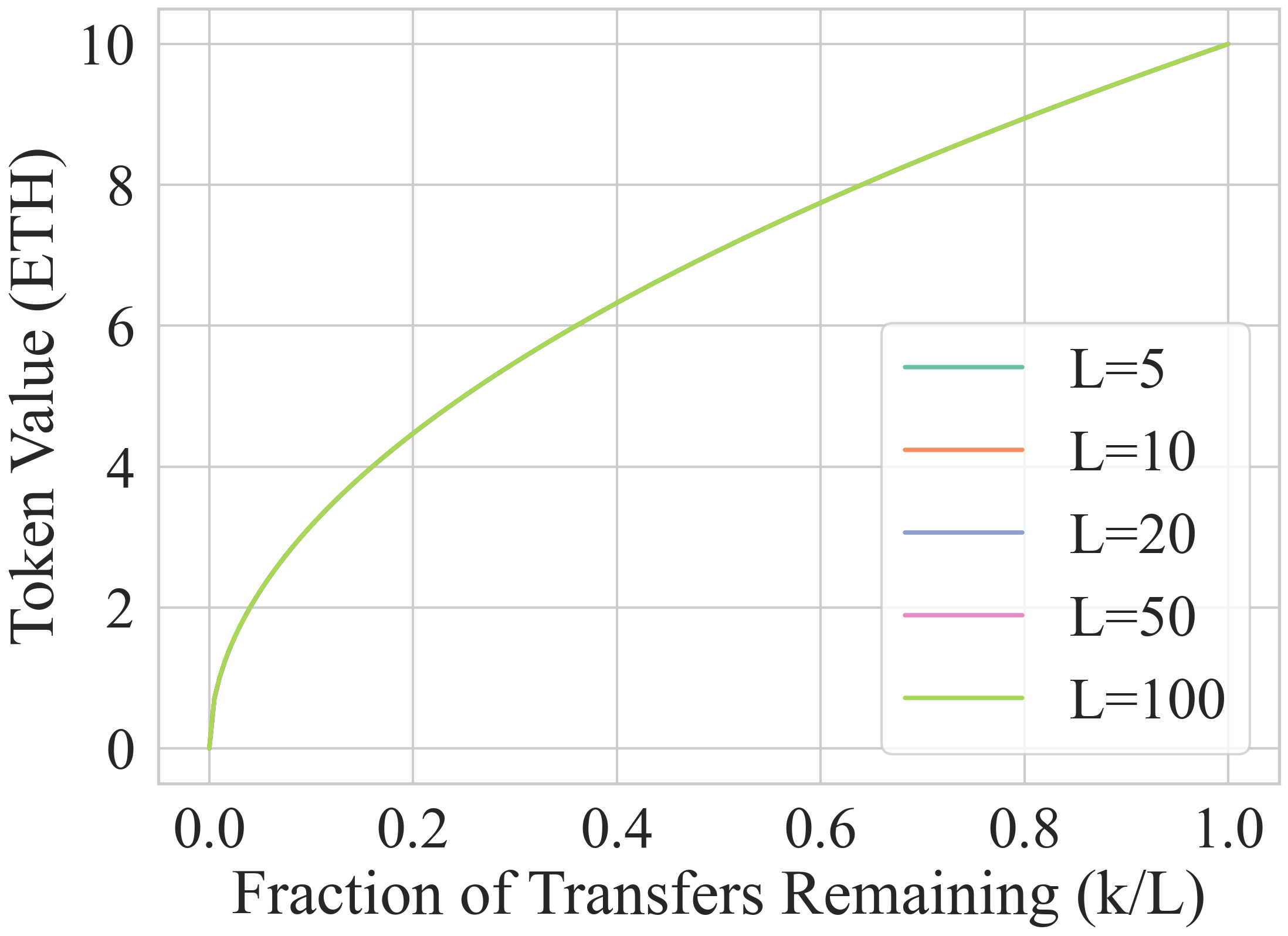}
        \caption{Concave model across limits.}
        \label{fig:valuation_b}
    \end{subfigure}
    \caption{Token value as a function of remaining transfers under four mobility premium models ($L=20$), and concave-model behavior across different transfer limits.}
    \label{fig:valuation}
\end{figure}

Table~\ref{tab:valuation} and Fig.\ref{fig:valuation} compare valuation dynamics across mobility premium models. The concave model ($\gamma=0.5$) yields the slowest value decay: at the midpoint ($k/L=0.5$), tokens retain 70.7\% of their base value, whereas the convex model falls to 25\%. Economically, the concave profile aligns with an \emph{option-value interpretation}, where each remaining transfer represents the opportunity to sell in the future, producing diminishing marginal value loss.

By contrast, the threshold model introduces a ``cliff effect,'' in which value drops sharply once remaining mobility falls below $\tau=0.2$. Such discontinuities may induce anticipatory selling as holders approach the boundary, potentially generating endogenous liquidity shocks. The convex model ($\gamma=2$) instead captures strong ``first-owner premium'' dynamics, where perceived freshness dominates valuation and early transfers incur substantial depreciation, consistent with luxury or original-edition assets.

\smallskip
\begin{center}
\fbox{%
\begin{minipage}{0.9\linewidth}
\textbf{Takeaway 2: The mobility premium.}
Each remaining transfer retains measurable value. Under the concave model ($\gamma{=}0.5$), a token preserves about 70.7\% of its base value halfway through its transfer budget, making successive transfers progressively more costly.
\end{minipage}}
\vspace{5pt}
\end{center}

\subsection{Marginal Mobility Cost}

% \begin{table}[t]
% \centering
% \caption{Marginal Mobility Cost as \% of $V_{\text{base}}$ (Concave Model, Selected Values).}
% \label{tab:marginal_cost}
% \begin{tabular}{c|cccc}
% \toprule
% \textbf{Remaining} & \textbf{$L=5$} & \textbf{$L=10$} & \textbf{$L=20$} & \textbf{$L=50$} \\
% \midrule
% First transfer    & 10.6\% & 5.1\%  & 2.5\%  & 1.0\% \\
% Mid-point         & 14.2\% & 10.1\% & 7.1\%  & 4.5\% \\
% Last transfer     & 44.7\% & 31.6\% & 22.4\% & 14.1\% \\
% \bottomrule
% \end{tabular}
% \end{table}

\begin{figure}[t]
    \centering
    \includegraphics[width=0.9\columnwidth]{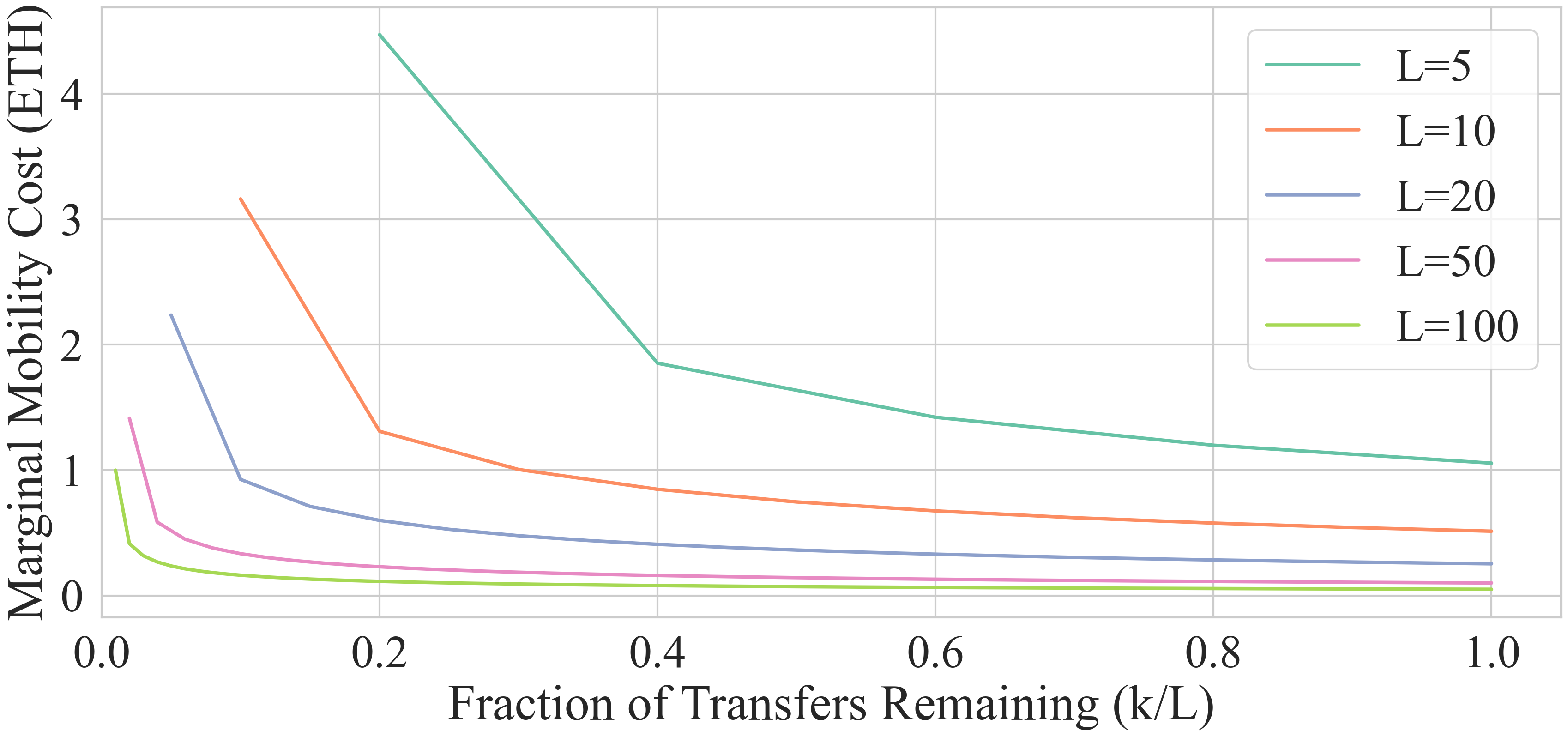}
    \caption{Marginal cost of successive transfers under the concave valuation model. Costs rise as remaining transfers decrease, with smaller limits amplifying the effect.}
    \label{fig:marginal_cost}
\end{figure}

Table~\ref{tab:marginal_cost} quantifies how marginal mobility costs evolve under the concave valuation model. Early transfers incur relatively small value loss, while later transfers become progressively expensive as mobility scarcity increases. This nonlinear escalation reflects diminishing remaining optionality: consuming one of the final transfers removes a disproportionately large fraction of future resale flexibility.

The effect strengthens as transfer limits decrease. For example, under $L=5$, the final transfer destroys 44.7\% of the base value, creating a strong endogenous deterrent against low-value or speculative transfers. Larger limits smooth the cost curve, but the increasing marginal penalty persists across all configurations, as illustrated in Fig.\ref{fig:marginal_cost}.

Economically, this mechanism converts mobility into a scarce resource whose consumption carries increasing opportunity cost. Rather than prohibiting transfers outright, the model induces self-regulation: rational holders are incentivized to reserve remaining transfers for higher-value ownership changes, aligning market behavior with long-term asset utility.

\subsection{Wash Trading Deterrence}

% \begin{table}[t]
% \centering
% \caption{Wash Trading Profitability in Cap-Aware Market (30\% inflation, $V_{\text{base}}=10$ ETH). Profit in ETH.}
% \label{tab:wash_profit}
% \begin{tabular}{cc|c|cc|cc}
% \toprule
% \multirow{2}{*}{\textbf{$L$}} & \multirow{2}{*}{\textbf{$n$}} & \textbf{No-Cap} & \textbf{Fair Val.} & \textbf{Max Sell} & \textbf{Cap} & \textbf{Det.?} \\
%  & & \textbf{Profit} & \textbf{After} & \textbf{(Cap)} & \textbf{Profit} & \\
% \midrule
% 5  & 1  & 3.00 & 8.94 & 11.63 & 1.62  & No \\
% 5  & 3  & 2.99 & 6.33 & 8.22  & $-$1.79 & Yes \\
% 5  & 5  & 2.98 & 0.00 & 0.00  & $-$10.0 & Yes \\
% 10 & 3  & 2.99 & 8.37 & 10.88 & 0.86  & No \\
% 10 & 5  & 2.98 & 7.07 & 9.19  & $-$0.83 & Yes \\
% 10 & 10 & 2.95 & 0.00 & 0.00  & $-$10.1 & Yes \\
% 20 & 5  & 2.98 & 8.66 & 11.26 & 1.23  & No \\
% 20 & 9  & 2.96 & 7.42 & 9.64  & $-$0.41 & Yes \\
% 20 & 15 & 2.93 & 5.00 & 6.50  & $-$3.58 & Yes \\
% \bottomrule
% \end{tabular}
% \end{table}

\begin{table}[t]
\centering
\small
\renewcommand{\arraystretch}{1.15}
\setlength{\tabcolsep}{4pt}

\caption{{Wash Trading Profitability under Transfer Caps} (30\% inflation, $V_{\text{base}}=10$ ETH). Profit in ETH.}
\label{tab:wash_profit}
\resizebox{\linewidth}{!}{
\begin{tabular}{c| c| >{\columncolor{capWeak}} c| c c| c c}
\toprule
\multirow{2}{*}{\textbf{$L$}} &
\multirow{2}{*}{\textbf{$n$}} &
\multicolumn{1}{c|}{\textbf{Baseline}} &
\multicolumn{2}{c|}{\textbf{Cap-aware valuation}} &
\multicolumn{2}{c}{\textbf{Outcome}} \\

\cmidrule(lr){3-3}
\cmidrule(lr){4-5}
\cmidrule(lr){6-7}

 & & \textbf{No-cap profit}
 & \textbf{Fair value}
 & \textbf{Max sell}
 & \textbf{Cap profit}
 & \textbf{Det.?} \\
\midrule

% \rowcolor{rowlabel}
5  & 1  & 3.00 & 8.94 & 11.63 & \cellcolor{capMedium}1.62  & No \\
   & 3  & 2.99 & 6.33 & 8.22  & \cellcolor{capStrong}$-$1.79 & Yes \\
   & 5  & 2.98 & 0.00 & 0.00  & \cellcolor{capStrong}$-$10.0 & Yes \\

\midrule

% \rowcolor{rowlabel}
10 & 3  & 2.99 & 8.37 & 10.88 & \cellcolor{capMedium}0.86  & No \\
   & 5  & 2.98 & 7.07 & 9.19  & \cellcolor{capStrong}$-$0.83 & Yes \\
   & 10 & 2.95 & 0.00 & 0.00  & \cellcolor{capStrong}$-$10.1 & Yes \\

\midrule

% \rowcolor{rowlabel}
20 & 5  & 2.98 & 8.66 & 11.26 & \cellcolor{capMedium}1.23  & No \\
   & 9  & 2.96 & 7.42 & 9.64  & \cellcolor{capStrong}$-$0.41 & Yes \\
   & 15 & 2.93 & 5.00 & 6.50  & \cellcolor{capStrong}$-$3.58 & Yes \\

\bottomrule
\end{tabular}
}
\end{table}

\begin{table}[t]
\centering
\caption{Maximum Leverage by Transfer Limit 
(LTV = 70\%, $V_0 = 10$ ETH).}
\label{tab:leverage}

\renewcommand{\arraystretch}{1.18}
\setlength{\tabcolsep}{6pt}

\begin{tabular}{c|c|c|c|c}
\toprule

\textbf{$L$} &
\textbf{Max Depth} &
\textbf{Max Exposure} &
\textbf{Eff. Leverage} &
\textbf{Reduction} \\
\midrule

\rowcolor{capStrong}
4   & 2  & 21.90 & \textbf{2.19$\times$} & 34.2\% \\

6   & 3  & 25.33 & 2.53$\times$ & 24.0\% \\

\rowcolor{capMedium}
10  & 5  & 29.41 & 2.94$\times$ & 11.7\% \\

20  & 10 & 32.67 & 3.27$\times$ & 1.9\% \\

\rowcolor{capWeak}
50  & 25 & 33.33 & 3.33$\times$ & 0.0\% \\

\bottomrule
\end{tabular}
\end{table}

\begin{figure}[t]
    \centering
    \begin{subfigure}[b]{0.48\columnwidth}
        \centering
        \includegraphics[width=\textwidth]{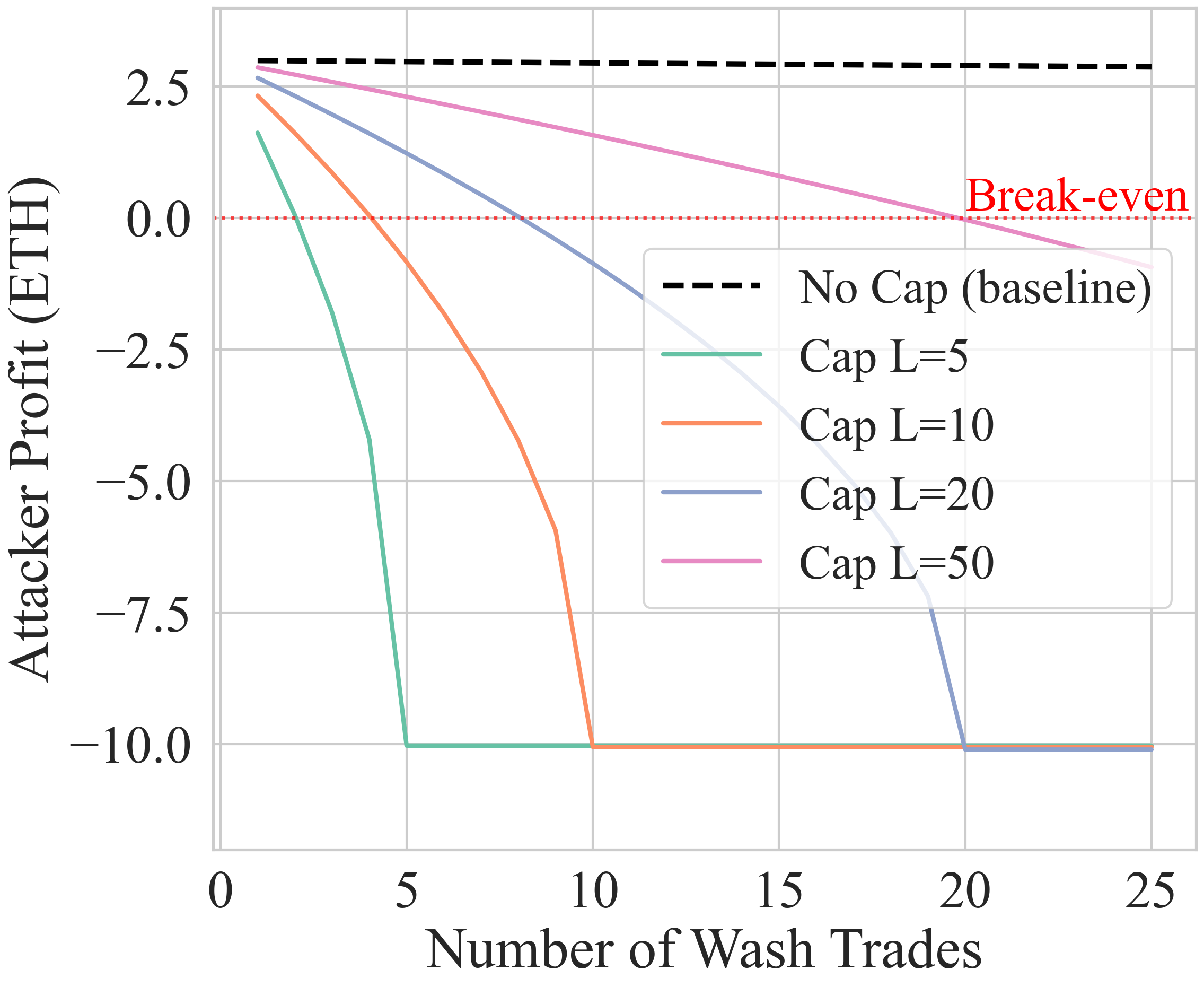}
        \caption{Profit vs.\ wash count.}
        \label{fig:wash_deterrence_a}
    \end{subfigure}
    \hfill
    \begin{subfigure}[b]{0.48\columnwidth}
        \centering
        \includegraphics[width=\textwidth]{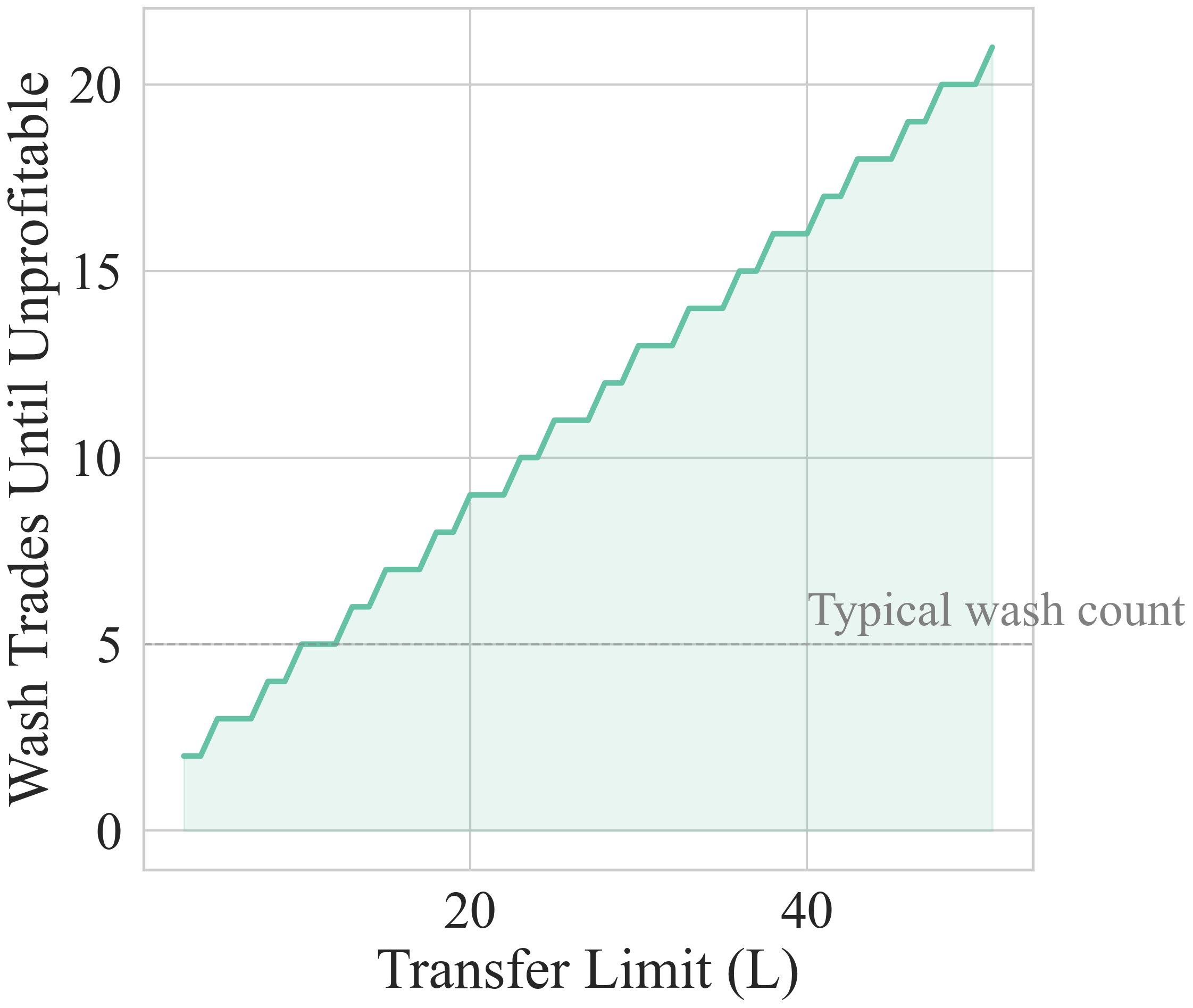}
        \caption{Break-even by transfer limit.}
        \label{fig:wash_deterrence_b}
    \end{subfigure}
    \caption{{Wash trading deterrence.} The dashed line marks zero profit. Break-even wash count increases with transfer limit $L$.}
    \label{fig:wash_deterrence}
\end{figure}

\begin{figure}[t]
    \centering
    \begin{subfigure}[b]{0.48\columnwidth}
        \centering
        \includegraphics[width=\textwidth]{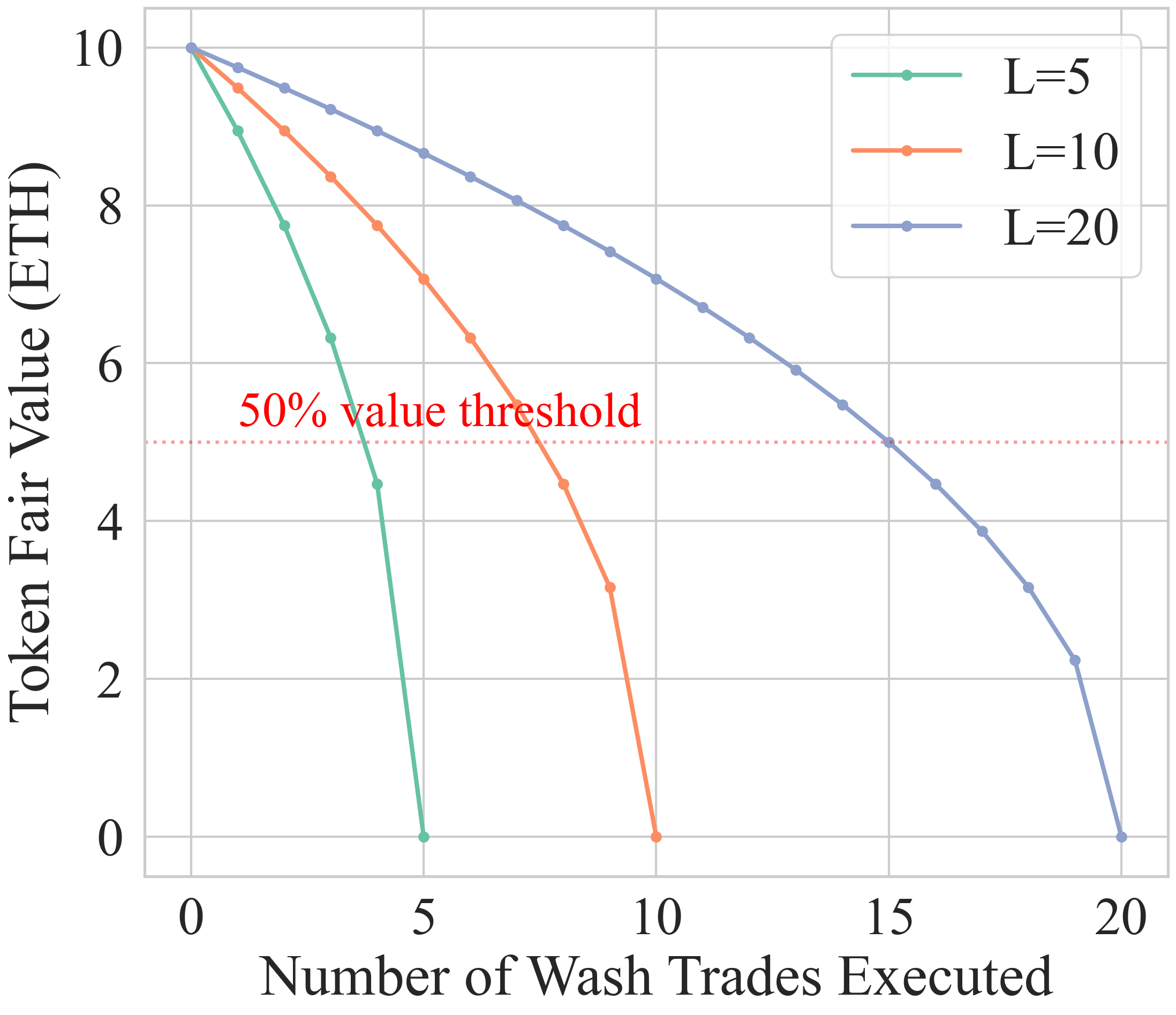}
        \caption{Value degradation by cap.}
        \label{fig:value_degrade_a}
    \end{subfigure}
    \hfill
    \begin{subfigure}[b]{0.48\columnwidth}
        \centering
        \includegraphics[width=\textwidth]{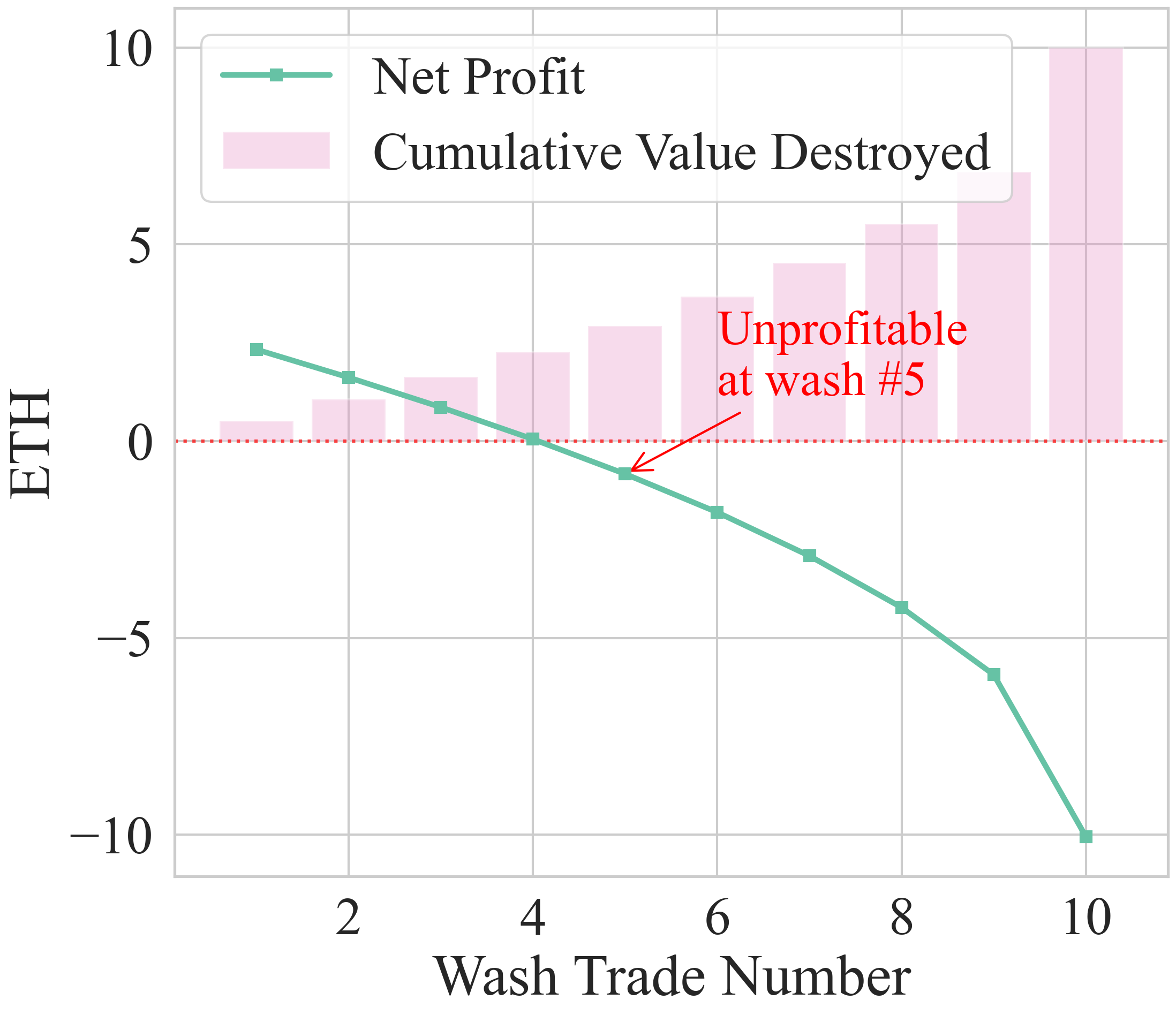}
        \caption{Profit trajectory ($L=10$).}
        \label{fig:value_degrade_b}
    \end{subfigure}
    \caption{{Token value degradation during wash trading.} Each wash trade permanently reduces token value by consuming transfer budget.}
    \label{fig:value_degrade}
\end{figure}

Table~\ref{tab:wash_profit} and Fig.\ref{fig:wash_deterrence} show that wash trading under transfer caps becomes inherently self-limiting. Each artificial trade consumes part of the finite mobility budget, permanently lowering the token's fair value. In a cap-aware market where buyers observe remaining transfer counts, the achievable resale price declines after every wash cycle.

For example, when $L=10$, wash-trading profitability becomes negative after five trades ($\Pi=-0.83$ ETH), whereas for $L=5$ the break-even point occurs after only three trades. As shown in Fig.\ref{fig:value_degrade_b}, profit crosses zero at wash~\#5 and decreases monotonically thereafter. This behavior arises from cumulative value degradation associated with transfer consumption, illustrated in Fig.\ref{fig:value_degrade_a}.

A key asymmetry drives this deterrence. The attacker alone bears irreversible value destruction, whereas honest holders preserve their remaining mobility budgets. Unlike traditional wash trading, where costs are largely limited to transaction fees, ERC-7634 converts manipulation into permanent asset depreciation. The deterrent therefore operates at the protocol level: economic incentives suppress repeated manipulation~\cite{wang2024cryptocurrency}.

\smallskip
\begin{center}
\fbox{%
\begin{minipage}{0.9\linewidth}
\textbf{Takeaway 3: Self-defeating wash trading.}
Transfer caps transform wash trading into an economically self-defeating strategy. Each fake trade irreversibly destroys value by consuming finite mobility.
\end{minipage}}
\vspace{5pt}
\end{center}

\subsection{Recursive Leverage Control}

% \begin{table}[t]
% \centering
% \caption{Maximum Leverage by Transfer Limit (LTV = 70\%, $V_0 = 10$ ETH)}
% \label{tab:leverage}
% \begin{tabular}{c|cc|cc}
% \toprule
% \textbf{$L$} & \textbf{Max Depth} & \textbf{Max Exposure} & \textbf{Eff. Leverage} & \textbf{Reduction} \\
% \midrule
% 4   & 2  & 21.90 & 2.19$\times$ & 34.2\% \\
% 6   & 3  & 25.33 & 2.53$\times$ & 24.0\% \\
% 10  & 5  & 29.41 & 2.94$\times$ & 11.7\% \\
% 20  & 10 & 32.67 & 3.27$\times$ & 1.9\% \\
% 50  & 25 & 33.33 & 3.33$\times$ & 0.0\% \\
% \bottomrule
% \end{tabular}
% \end{table}

\begin{figure}[t]
    \centering
    \begin{subfigure}[b]{0.48\columnwidth}
        \centering
        \includegraphics[width=\textwidth]{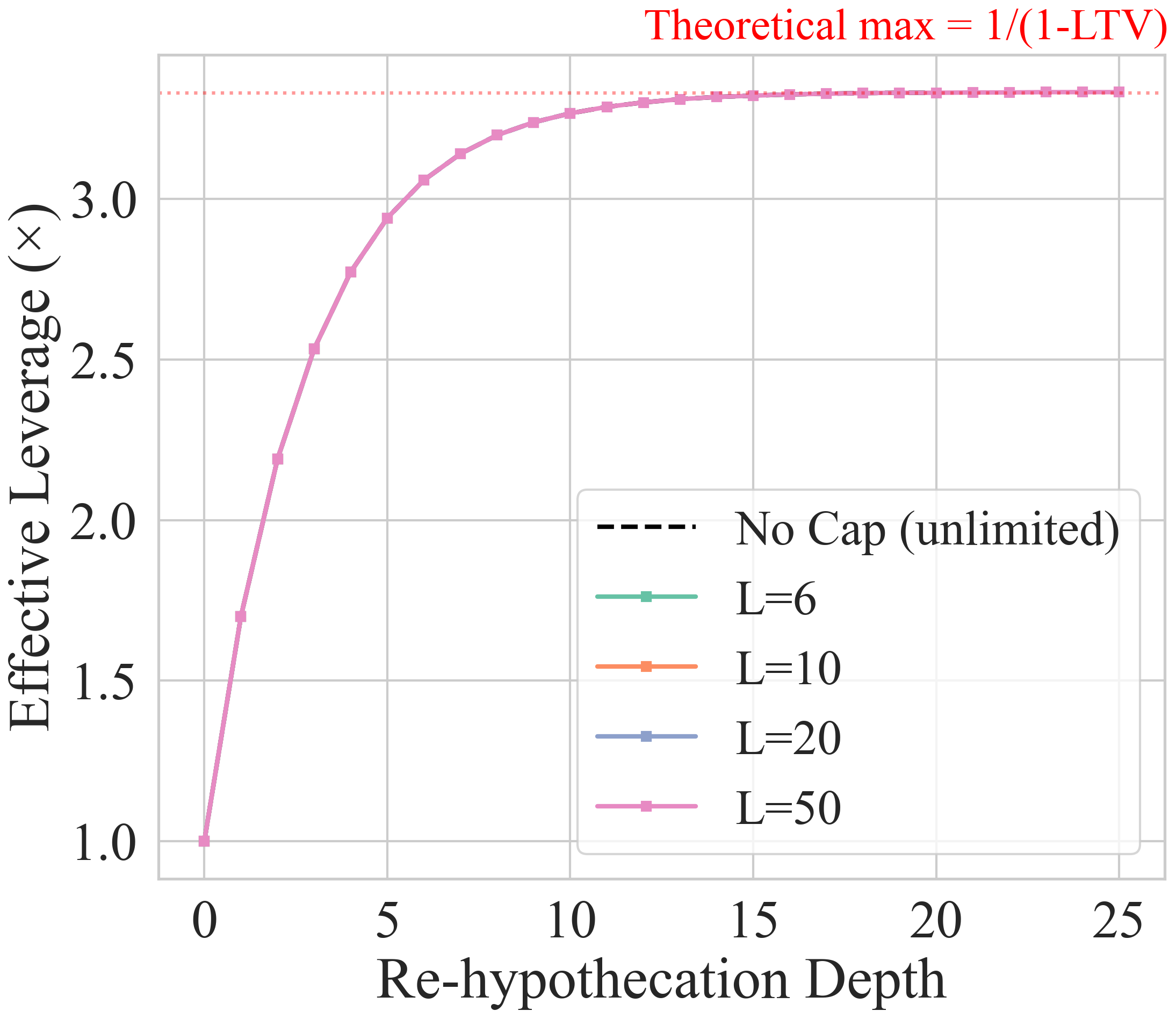}
        \caption{Leverage vs.\ chain depth.}
        \label{fig:leverage_a}
    \end{subfigure}
    \hfill
    \begin{subfigure}[b]{0.48\columnwidth}
        \centering
        \includegraphics[width=\textwidth]{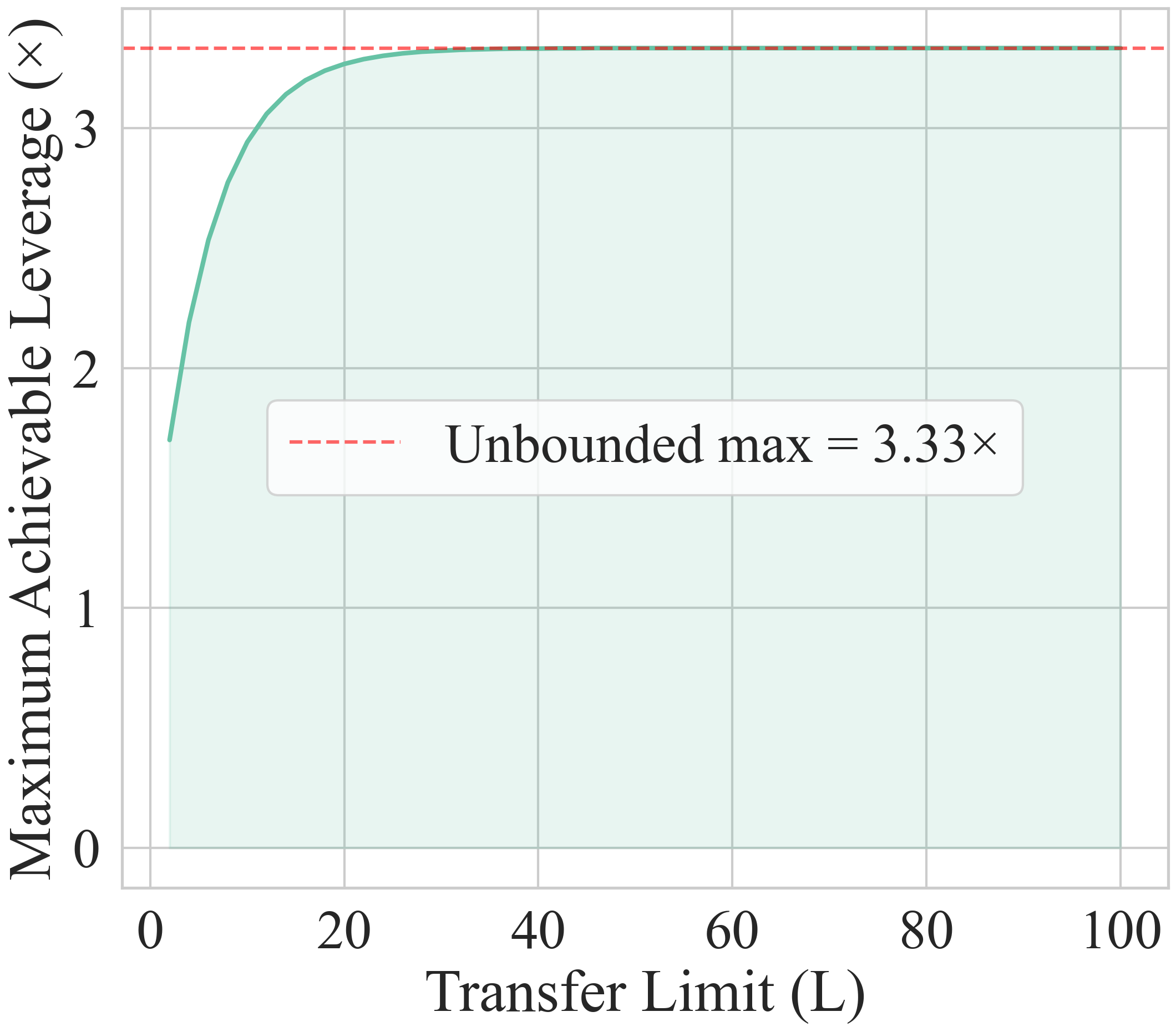}
        \caption{Cap-bounded maxi leverage.}
        \label{fig:leverage_b}
    \end{subfigure}
    \caption{Effective leverage under bounded transfers. Leverage growth with re-hypothecation depth and maximum achievable leverage as a function of transfer limit $L$. The dashed line indicates the theoretical unbounded limit.}
    \label{fig:leverage}
\end{figure}

\begin{figure}[t]
    \centering
    \includegraphics[width=0.9\columnwidth]{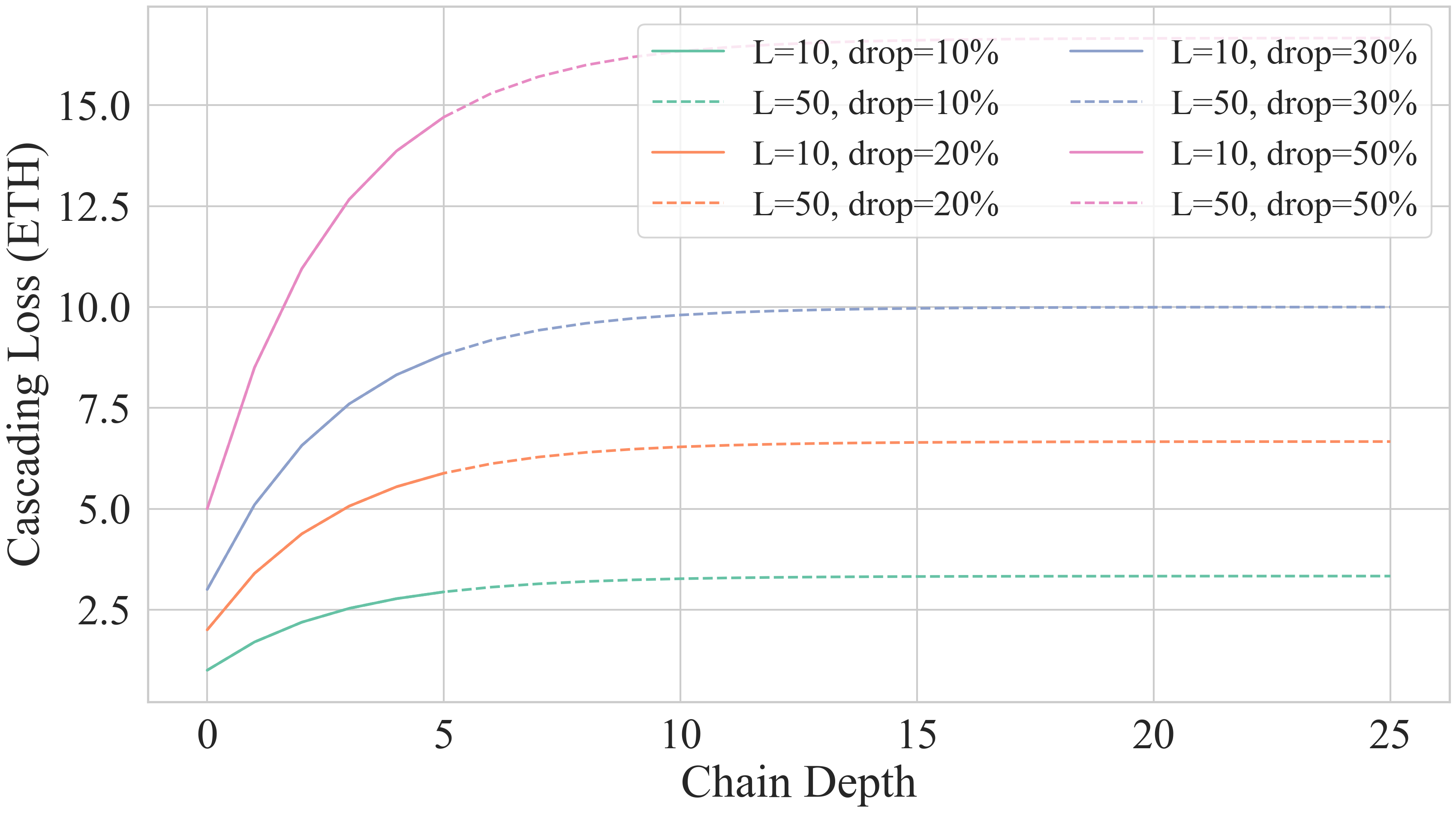}
    \caption{{Cascading losses under leverage chains.} Transfer caps ($L = 10$) shorten leverage chains and reduce loss propagation compared to uncapped scenarios ($L = 50$).}
    \label{fig:systemic_risk}
\end{figure}

\begin{figure}[t]
    \centering
    \begin{subfigure}[b]{0.48\columnwidth}
        \centering
        \includegraphics[width=\textwidth]{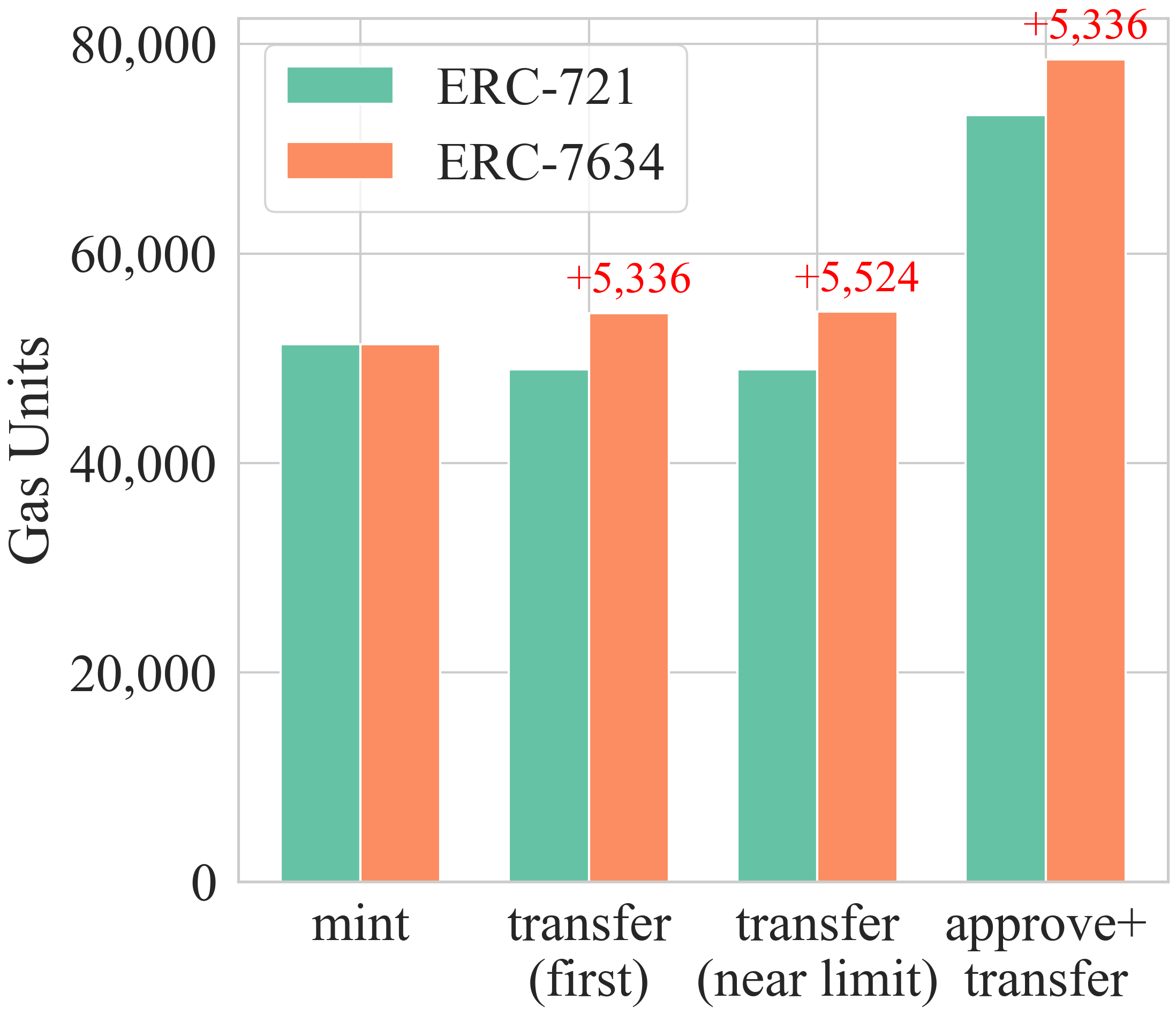}
        \caption{Absolute gas costs.}
        \label{fig:gas_a}
    \end{subfigure}
    \hfill
    \begin{subfigure}[b]{0.48\columnwidth}
        \centering
        \includegraphics[width=\textwidth]{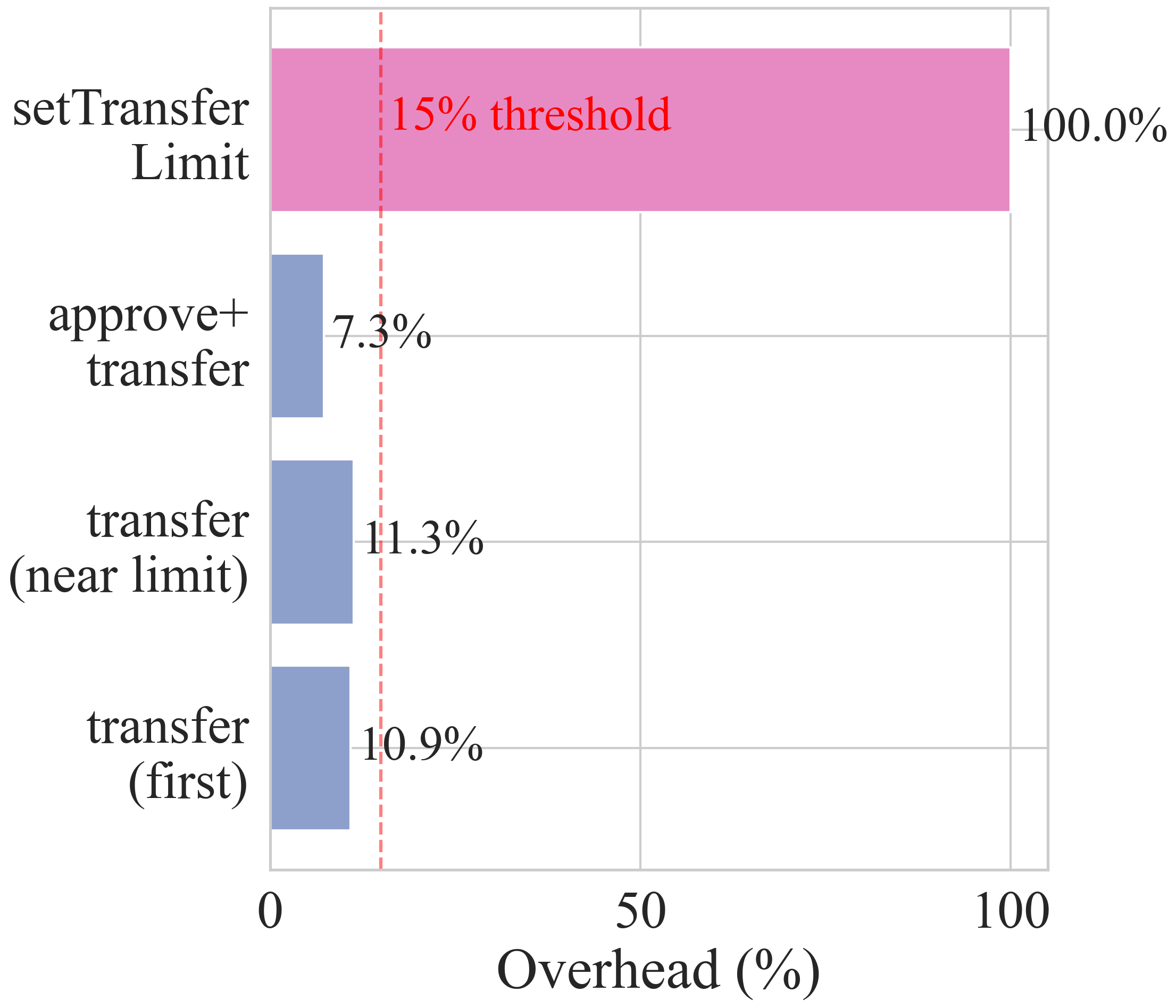}
        \caption{Percentage overhead.}
        \label{fig:gas_b}
    \end{subfigure}
    \caption{Gas costs. ERC-7634 introduces modest overhead relative to the ERC-721 baseline across core operations.}
    \label{fig:gas}
\end{figure}

Table~\ref{tab:leverage} shows that transfer caps impose a direct structural bound on recursive leverage. Smaller limits substantially reduce achievable exposure: at $L=4$, effective leverage is capped at $2.19\times$, a 34.2\% reduction from the theoretical maximum. As $L$ increases, leverage approaches the unbounded limit, confirming that mobility constraints act as a truncation mechanism on geometric leverage growth.

The mechanism parallels reserve requirements in traditional finance. Just as reserve ratios limit the money multiplier, transfer caps restrict the ``collateral multiplier'' by bounding re-hypothecation depth. Notably, intermediate limits provide the strongest practical trade-off. For instance, $L=6$ achieves a 24.0\% leverage reduction while allowing three collateral cycles, preserving ordinary trading flexibility while constraining excessive leverage formation.

Fig.\ref{fig:leverage} visualizes how leverage accumulation saturates earlier under capped regimes. Fig.\ref{fig:systemic_risk} further demonstrates the implication: shorter leverage chains produce materially smaller cascading losses under adverse price shocks. Under a 30\% price decline, capped systems ($L=10$) limit both cascade depth and aggregate losses relative to uncapped markets.

\smallskip
\begin{center}
\fbox{%
\begin{minipage}{0.9\linewidth}
\textbf{Takeaway 4: Leverage bounded by design.}
Transfer caps constrain restaking depth. With $L{=}10$, maximum leverage decreases from $3.33\times$ (unbounded) to $2.94\times$, achieving an 11.7\% reduction without modifying lending protocols or introducing external controls.
\end{minipage}}
\vspace{5pt}
\end{center}

% ERC-7634 introduces minimal runtime overhead by adding two lightweight operations to the standard ERC-721 transfer path: (i) a storage read to verify that the current transfer count does not exceed the configured limit in \texttt{\_beforeTokenTransfer}, and (ii) a storage write with event emission to increment the counter in \texttt{\_afterTokenTransfer}. The measured gas overhead is summarized below:

% \begin{itemize}
%     \item \textit{Standard transfer overhead:} approximately 5{,}336 gas, corresponding to a $\sim$10.9\% increase over the ERC-721 baseline (48{,}947 gas).
    
%     \item \textit{Near-limit transfer:} approximately 5{,}524 gas ($\sim$11.3\%), slightly higher due to additional boundary-condition evaluation when the transfer count approaches the cap.
    
%     \item \textit{One-time configuration cost:} invoking \texttt{setTransferLimit} requires $\sim$23{,}496 gas, which is amortized over the lifetime of the token and independent of subsequent transfers.
    
%     \item \textit{View operations:} read-only queries such as \texttt{transferCountOf} and \texttt{transferLimitOf} consume $\sim$2{,}400 gas each and are effectively negligible for off-chain indexing and analytics.
% \end{itemize}
\subsection{Gas Overhead Analysis}

% \begin{table}[t]
% \centering
% \caption{Gas Cost Comparison: ERC-721 vs.\ ERC-7634}
% \label{tab:gas}
% \begin{tabular}{c|ccc}
% \toprule
% \textbf{Operation} & \textbf{ERC-721} & \textbf{ERC-7634} & \textbf{Overhead} \\
% \midrule
% Mint                & 51{,}316 & 51{,}316 & 0.0\% \\
% Mint + setLimit     & --       & 74{,}812 & -- \\
% Transfer (first)    & 48{,}947 & 54{,}283 & 10.9\% \\
% Transfer (near cap) & 48{,}947 & 54{,}471 & 11.3\% \\
% Approve + transfer  & 73{,}221 & 78{,}557 & 7.3\% \\
% setTransferLimit    & --       & 23{,}496 & -- \\
% \bottomrule
% \end{tabular}
% \end{table}

Table~\ref{tab:gas} and Fig.\ref{fig:gas} show that ERC-7634 incurs only moderate execution overhead. Standard transfers increase by approximately 10.9\%, primarily due to one additional \texttt{SLOAD} operation for limit verification and one \texttt{SSTORE} operation for counter updates. Transfers executed near the cap exhibit only marginally higher cost (11.3\%), indicating stable runtime behavior independent of transfer state.

The one-time \texttt{setTransferLimit} operation introduces an initialization cost of 23{,}496 gas, which is amortized over the token lifecycle and does not affect subsequent transfers. Importantly, mint operations remain unchanged, preserving compatibility with existing minting pipelines and batch issuance workflows.

\subsection{Security Cost Analysis}

\begin{figure}[t]
    \centering
    \begin{subfigure}[b]{0.48\columnwidth}
        \centering
        \includegraphics[width=\textwidth]{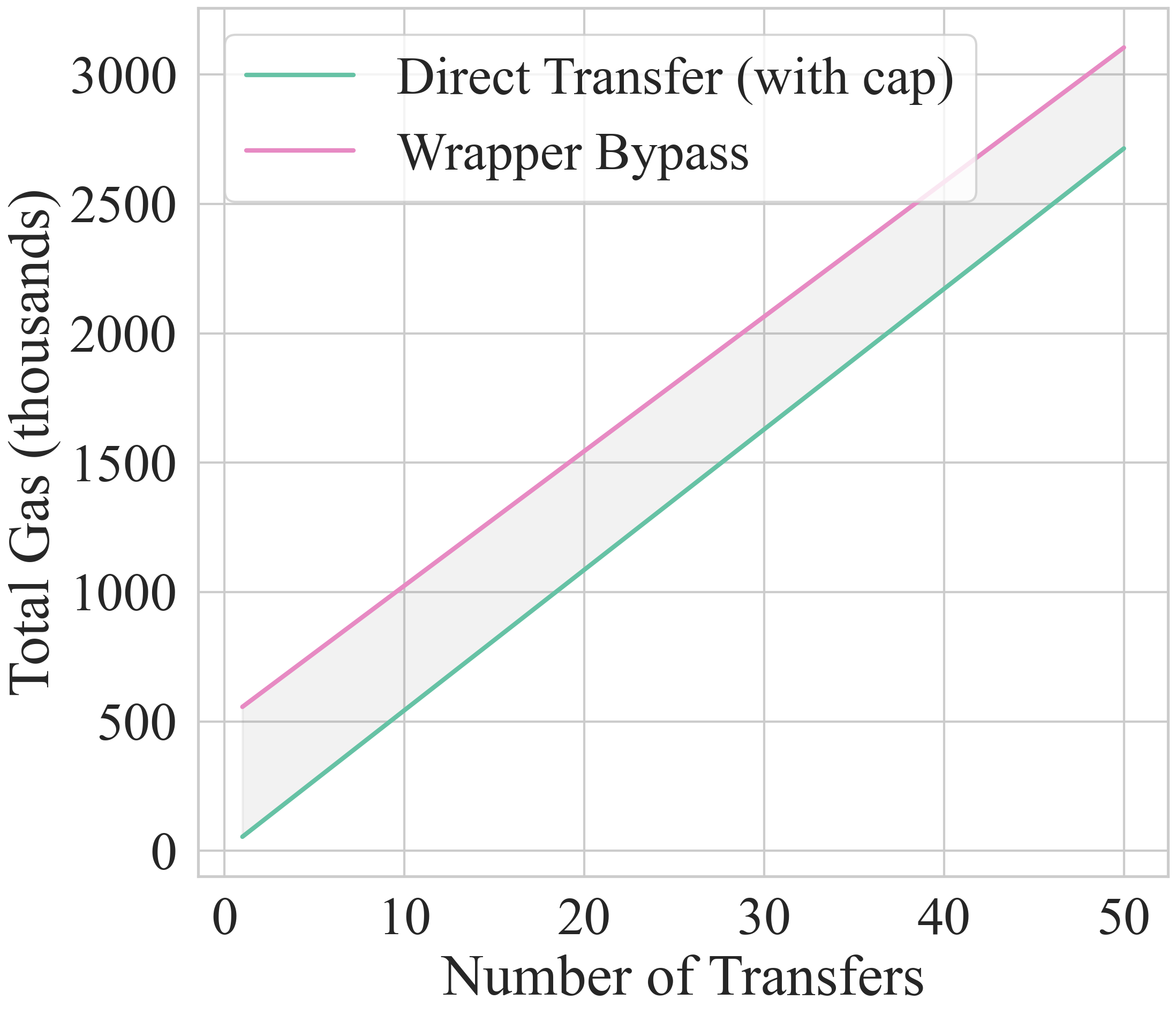}
        \caption{Direct vs wrapper gas costs.}
        \label{fig:security_a}
    \end{subfigure}
    \hfill
    \begin{subfigure}[b]{0.48\columnwidth}
        \centering
        \includegraphics[width=\textwidth]{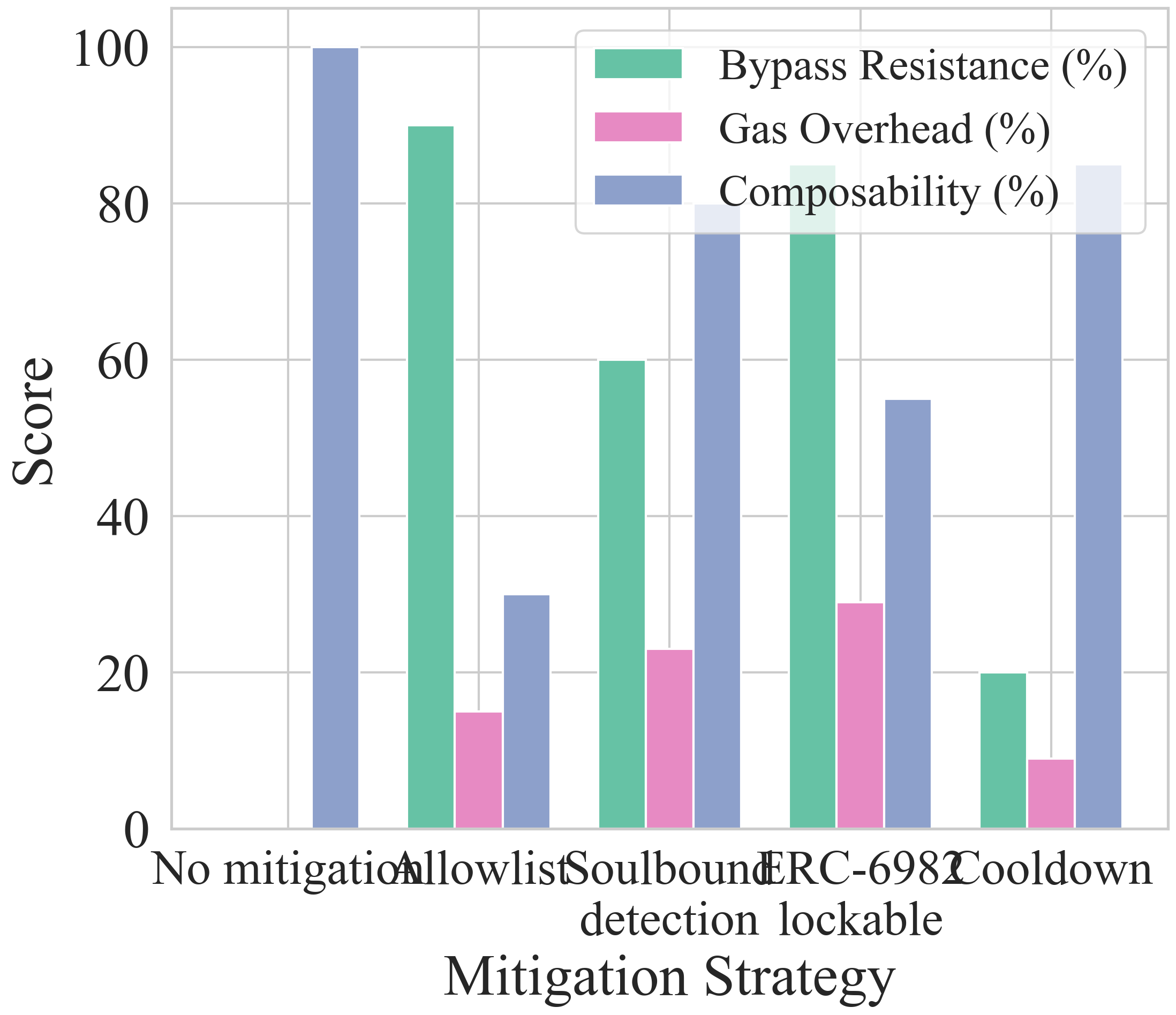}
        \caption{Mitigation strategy trade-offs.}
        \label{fig:security_b}
    \end{subfigure}
    \caption{Security cost. The wrapper bypass breaks even at $N \approx 221$ transfers. Trade-offs between bypass resistance, gas overhead, and composability across five mitigation strategies.}
    \label{fig:security}
\end{figure}

Fig.\ref{fig:security_a} shows that the wrapper bypass reaches break-even at approximately $N = 221$ transfers, far exceeding typical NFT lifetimes. This threshold is parameter-dependent and is computed under our default gas assumptions and measured per-transfer wrapper overhead in Fig.\ref{fig:security_a}. For comparison, the median transfer count across collections is only 1--2 (Table~\ref{tab:transfer_stats}), while even the P99 for gaming items, the most actively traded category, reaches 304 transfers. Thus, bypass becomes economically rational only for extreme tail cases. In addition, wrapper tokens lose native ERC-7634 metadata: wrapped assets cannot expose \texttt{transferCountOf} or \texttt{transferLimitOf}, weakening credibility in cap-aware marketplaces where buyers rely on transparent transfer budgets. Fig.\ref{fig:security_b} summarizes mitigation trade-offs, showing that ERC-6982 lockable integration provides the strongest balance between bypass resistance (85\%) and composability (55\%).

\smallskip
\begin{center}
\fbox{%
\begin{minipage}{0.9\linewidth}
\textbf{Takeaway 5: practical cost profile.}
ERC-7634 adds less than 11\% gas overhead per transfer while making wrapper bypass often economically unattractive under our baseline assumptions: the break-even point of ${\sim}221$ transfers lies well beyond typical NFT lifetimes, limiting feasibility to rare edge cases.
\end{minipage}
}
\vspace{5pt}
\end{center}
% ══════════════════════════════════════════════════════════════════════════════
%  7. DESIGN SPACE TAXONOMY
% ══════════════════════════════════════════════════════════════════════════════

\section{Design Space Taxonomy}
\label{sec:taxonomy}

\begin{table}[b]
\centering
\caption{Gas Cost Comparison: ERC-721 vs.\ ERC-7634}
\label{tab:gas}

\renewcommand{\arraystretch}{1.18}
\setlength{\tabcolsep}{8pt}

\begin{tabular}{c|cc|c}
\toprule

\textbf{Operation} & \textbf{ERC-721} & \textbf{ERC-7634} & \textbf{Overhead} \\
\midrule

Mint                & 51{,}316 & 51{,}316 & 0.0\% \\

Mint + setLimit     & \textit{--} & 74{,}812 & \textit{N/A} \\

\rowcolor{capStrong}
Transfer (first)    & 48{,}947 & 54{,}283 & \textbf{10.9\%} \\

\rowcolor{capStrong}
Transfer (near cap) & 48{,}947 & 54{,}471 & \textbf{11.3\%} \\

Approve + transfer  & 73{,}221 & 78{,}557 & 7.3\% \\

setTransferLimit    & \textit{--} & 23{,}496 & \textit{N/A} \\

\bottomrule
\end{tabular}
\end{table}

\begin{table}[b]
\centering
\caption{Use Case Suitability (0--5 scale, 5 = ideal).}
\label{tab:use_cases}

\renewcommand{\arraystretch}{1.18}
\setlength{\tabcolsep}{5pt}

\begin{tabular}{c|c|c|c|c}
\toprule
\textbf{Use Case} & \multicolumn{1}{c}{\textbf{ERC-721}} & \multicolumn{1}{c}{\textbf{ERC-5192}} & \textbf{ERC-6982} & \textbf{ERC-7634} \\
\midrule
Identity        & 1 & 5 & 2 & 3 \\

DeFi collateral & \cellcolor{capMedium}5 &  0 & \cellcolor{capWeak}4 & \cellcolor{capWeak}\textbf{4} \\
RWA             & 3 & 1 & 3 & \cellcolor{capWeak} \textbf{4} \\
Loyalty         & 3 & \cellcolor{capWeak} 4 & 3 & \cellcolor{capWeak}\textbf{4} \\
Digital art     & \cellcolor{capMedium} 5 & 0 & 3 & \cellcolor{capWeak}\textbf{4} \\

Gaming items    & 3 & 0 & 3 & \cellcolor{capMedium}\textbf{5} \\
Event tickets   & 1 & 3 & 3 & \cellcolor{capMedium}\textbf{5} \\
Memberships     & 2 & \cellcolor{capWeak}4 & 3 & \cellcolor{capMedium}\textbf{5} \\

\bottomrule
\end{tabular}
\end{table}

\begin{figure}[t]
    \centering
    \begin{subfigure}[b]{0.48\columnwidth}
        \centering
        \includegraphics[width=\textwidth]{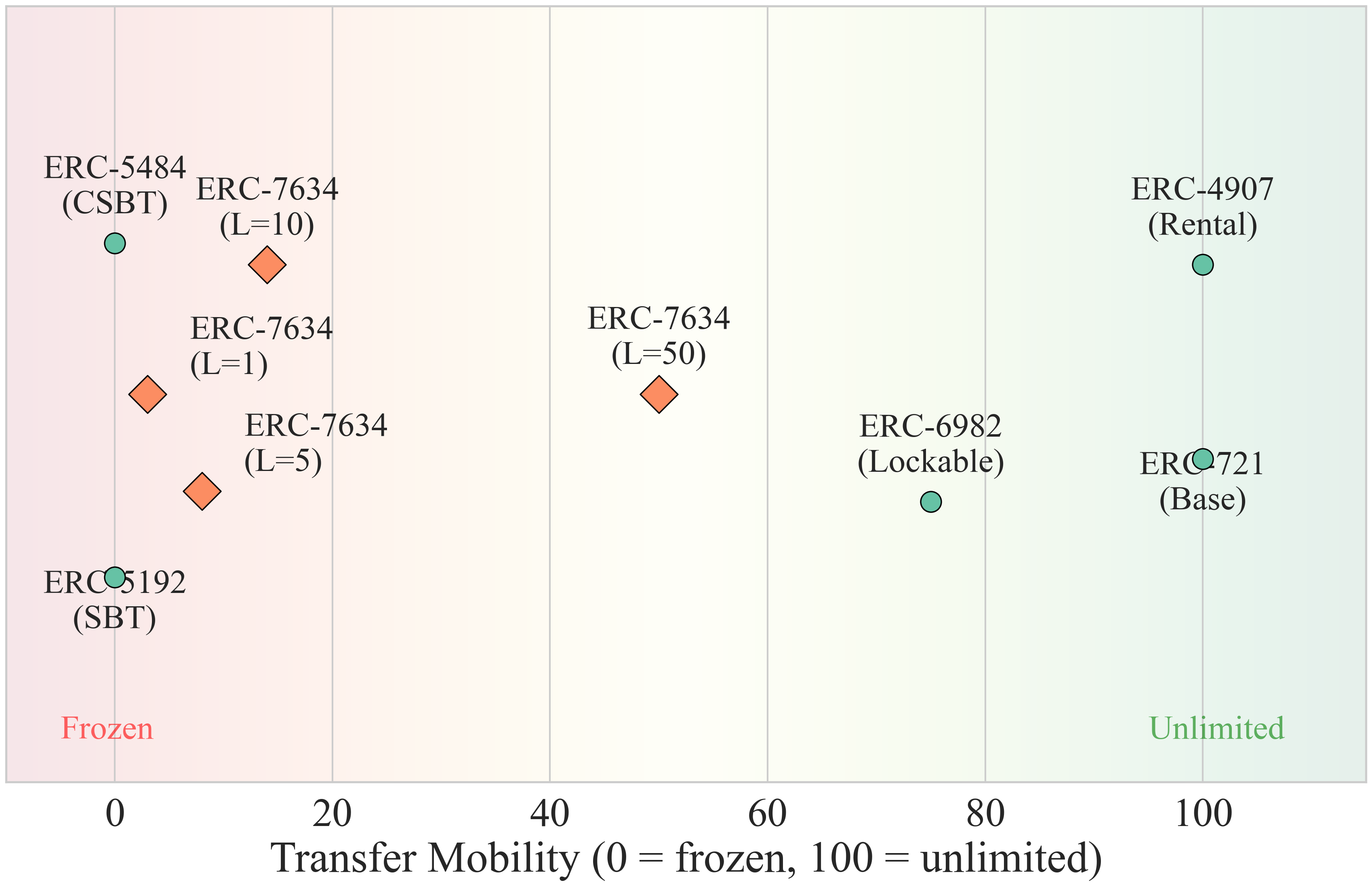}
        \caption{Token mobility spectrum.}
        \label{fig:design_space_a}
    \end{subfigure}
    \hfill
    \begin{subfigure}[b]{0.48\columnwidth}
        \centering
        \includegraphics[width=\textwidth]{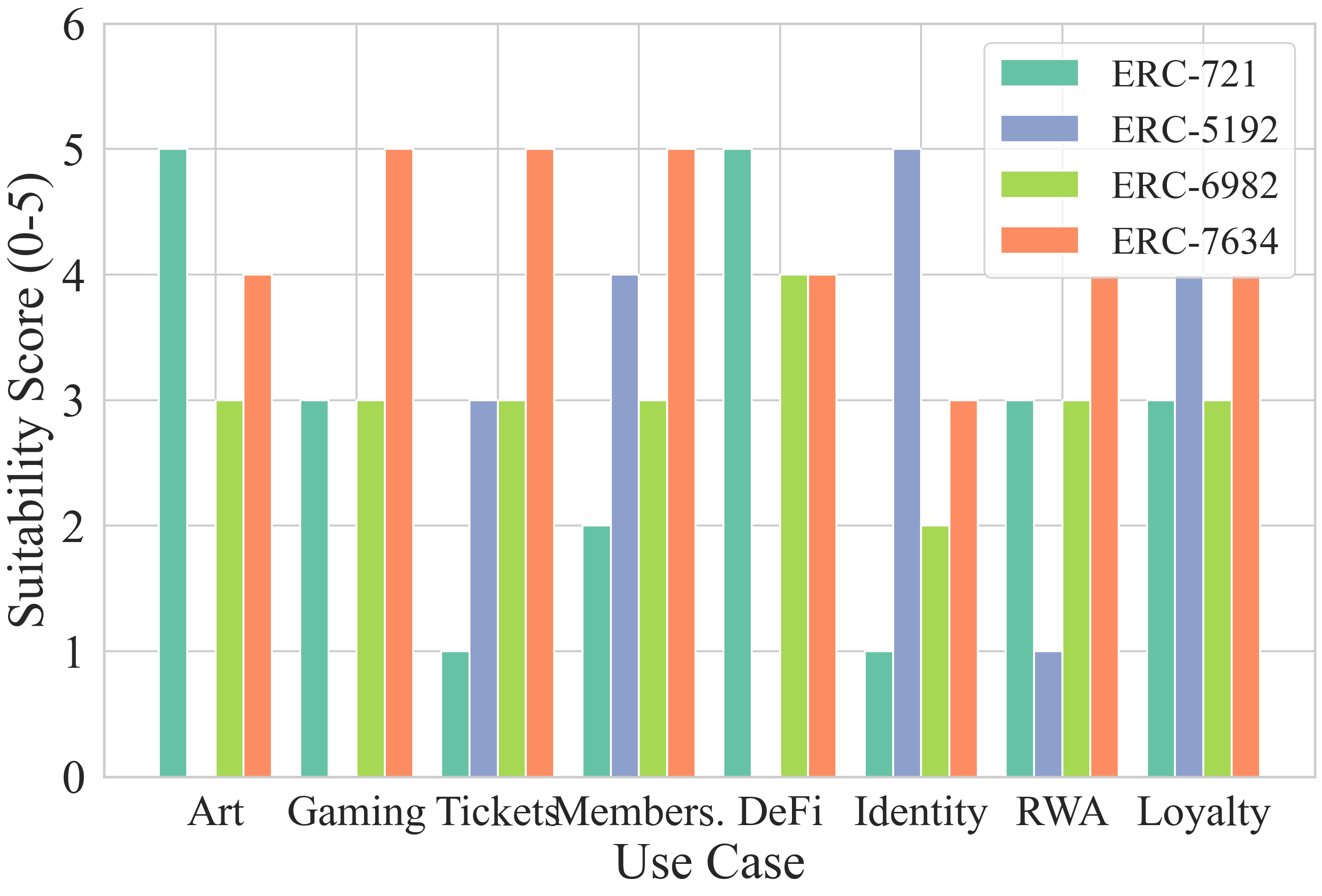}
        \caption{Use case suitability.}
        \label{fig:design_space_b}
    \end{subfigure}
    \caption{ERC-7634 (diamonds) bridges the gap between non-transferable SBTs and transferable ERC-721 tokens.}
    \label{fig:design_space}
\end{figure}

% \begin{figure}[t]
%     \centering
%     \includegraphics[width=0.85\linewidth]{figures/token-lifecycle/figure.png}
%     \caption{ERC-7634 token lifecycle: from minted (full budget) through active trading to a settled state, with optional extensions such as auto-burn (ERC-5679) or soulbound conversion (ERC-5192).}
%     \label{fig:lifecycle}
% \end{figure}

ERC-7634 is among the few standards that combine transfer-count awareness (Table~\ref{tab:taxonomy}), gradual value decay, broad DeFi compatibility, and medium-to-high anti-speculation capability.
Table~\ref{tab:use_cases} and Fig.\ref{fig:design_space} further show that ERC-7634 achieves the highest suitability for gaming items, event tickets, and memberships while remaining competitive for art, DeFi, and RWA scenarios. 

Unlike prior standards that treat mobility as a binary property, ERC-7634 introduces mobility as a continuous parameter controlled by the cap $L$. Small limits approximate soulbound behavior ($L=1$), moderate limits ($L=10$--$20$) enable controlled circulation, and large limits ($L \geq 50$) approach unrestricted transferability. This parameterized design allows a single standard to span the full applications in Table~\ref{tab:use_cases}, with $L$ acting as the primary configuration knob.

% Fig.\ref{fig:lifecycle} illustrates the lifecycle semantics enabled by bounded transfers: tokens evolve from minted (full budget) to active (partially consumed budget) and finally to settled (exhausted budget). Optional extensions further enable automatic burn or permanent binding, introducing a predictable temporal evolution absent in prior NFT standards and aligning incentives between issuers and holders.

% \smallskip
% \begin{center}
% \fbox{%
% \begin{minipage}{0.9\linewidth}
% \textbf{Takeaway 6: filling the mobility gap.}
% ERC-7634 uniquely combines transfer-count awareness, gradual value decay, full DeFi compatibility, and medium-to-high anti-speculation properties, achieving strong suitability for gaming, ticketing, and membership applications while remaining broadly applicable across NFT ecosystems.
% \end{minipage}
% }
% \vspace{5pt}
% \end{center}

% ══════════════════════════════════════════════════════════════════════════════
%  8. DISCUSSION
% ══════════════════════════════════════════════════════════════════════════════
% \clearpage

\section{Discussion}
\label{sec:discussion}

\subsection{Budget vs. Enforcement: An Honest Assessment}

ERC-7634 provides a \emph{budgeting} primitive rather than a strict \emph{enforcement} guarantee. As discussed in Section~\ref{sec:security}, wrapper bypass allows determined actors to circumvent transfer caps at the cost of wrapper deployment and reduced composability. We consider this trade-off acceptable for three reasons. First, the wrapper break-even point ($N \approx 221$) exceeds typical NFT transfer lifetimes by an order of magnitude. Second, wrapped tokens lose native ERC-7634 metadata, weakening their valuation in cap-aware marketplaces that rely on transparent transfer signals. Third, this model parallels real-world DRM systems: sufficiently effective for normal economic behavior without claiming absolute resistance to adversarial circumvention.

\subsection{Cap Selection}

Based on empirical transfer distributions, we propose practical cap-selection guidelines:

\begin{itemize}
    \item \textit{Memberships/licenses:} $L = 3\text{--}5$ (few legitimate resales; 94--97\% of tokens unaffected).
    \item \textit{Event tickets:} $L = 1\text{--}3$ (anti-scalping with limited resale flexibility).
    \item \textit{Art/collectibles:} $L = 10\text{--}20$ (supports secondary markets while bounding speculation; 85--97\% unaffected).
    \item \textit{Gaming items:} $L = 20\text{--}50$ (supports active trading while enabling lifecycle degradation).
    % \item \textit{DeFi collateral:} $L = 6\text{--}10$ (bounds leverage to approximately $2.5\text{--}2.9\times$).
\end{itemize}

\subsection{Post-Cap Token Destinations}
\label{sec:post_cap}

When a token exhausts its transfer budget ($k = L$), the protocol must define its subsequent state. ERC-7634 intentionally remains agnostic to post-cap behavior; however, its composability with existing standards enables four practical destination paths.

\smallskip
\noindent\textbf{Soulbound conversion (ERC-5192).} The token becomes permanently non-transferable at the current holder. This models credentials, achievements, or memberships that ``vest'' through use: after exhausting its allowed transfers, the NFT settles as a permanent proof of provenance tied to the final owner.

\smallskip
\noindent\textbf{Auto-burn (ERC-5679).} The token is destroyed once the cap is reached. This captures consumable assets such as event tickets, temporary licenses, or in-game items with finite durability, with the burn event serving as a clear on-chain lifecycle completion signal.

\smallskip
\noindent\textbf{Lock-and-release (ERC-6982).} The token becomes locked but may later be unlocked by governance or an oracle, optionally resetting mobility. This supports renewable subscriptions, seasonal passes, or condition-dependent assets whose transferability is restored under external events.

\smallskip
\noindent\textbf{Provenance freeze.} No companion standard is required; the token simply remains with its final owner. The resulting ownership history forms a complete and bounded provenance chain of exactly $L$ transfers, valuable for fine art and collectible assets where verified ownership lineage carries intrinsic value.

% \begin{table}[t]
% \centering
% \caption{Post-Cap Token Destination Paths}
% \label{tab:post_cap}
% \begin{tabular}{c|cc|c}
% \toprule
% \textbf{Path} & \textbf{Companion} & \textbf{Trigger} & \textbf{Use Case} \\
% \midrule
% Soulbound     & ERC-5192 & $k{=}L$          & Credentials \\
% Auto-burn     & ERC-5679 & $k{=}L$          & Tickets \\
% Lock-release  & ERC-6982 & $k{=}L$ + oracle & Subscriptions \\
% Provenance    & (none)   & $k{=}L$          & Art, collectibles \\
% \bottomrule
% \end{tabular}
% \end{table}

% requires: \usepackage{tabularx,booktabs,colortbl,xcolor}

\begin{table}[t]
\centering
% \caption{Post-Cap Token Destination Paths}
\caption{\textit{Post-Cap Token Destination Paths.} The table summarizes four representative behaviors available once a token exhausts its transfer budget ($k{=}L$).}
\label{tab:post_cap}

\footnotesize
\renewcommand{\arraystretch}{1.15}
\setlength{\tabcolsep}{5pt}

\begin{tabularx}{\linewidth}{c|c|c|>{\raggedright\arraybackslash}X}
\toprule
\multicolumn{1}{c}{\textbf{Path}} & \multicolumn{1}{c}{\textbf{Companion}} & \multicolumn{1}{c}{\textbf{Trigger}} & \multicolumn{1}{c}{\textbf{Use case}} \\
\midrule
\cellcolor{capMedium}Soulbound
& ERC-5192
& $k = L$
& Credentials/ID-binding. \\

\cellcolor{capMedium}Auto-burn
& ERC-5679
& $k = L$
& One-time consumables. \\

\cellcolor{capMedium}Lock-release
& ERC-6982
& $k =L$+oracle
& Subscriptions. \\

\cellcolor{capMedium}Provenance
& ---
& $k = L$
& Collectibles and arts. \\
\bottomrule
\end{tabularx}
\end{table}

Table~\ref{tab:post_cap} summarizes these destination paths, their companion standards, and representative use cases. Implementers select the desired behavior at deployment time, typically by extending the transfer enforcement hook to trigger the corresponding action once $k$ reaches $L$.

\subsection{Limitations}

\begin{packeditemize}
    \item \textit{Simulation fidelity.} Transfer distributions are synthetic and calibrated to published statistics; validation using real Ethereum on-chain data would strengthen external validity.
    \item \textit{Valuation assumptions.} The mobility premium models are analytical abstractions; empirical pricing data from transfer-bounded tokens is required to determine which functional form reflects real market behavior.
    \item \textit{Cross-chain transfers.} Transfer counts are chain-local. Bridging NFTs across chains may reset or bypass counts without cross-chain state synchronization.
    \item \textit{Limit mutability.} If \texttt{setTransferLimit} permits increases post-deployment, compromised or malicious owners could effectively remove caps; immutable or decrease-only limits may be preferable for high-stakes deployments.
\end{packeditemize}

% \subsection{Future Work}

% Several directions warrant further study: (i) empirical validation using Ethereum transfer data, (ii) formal verification of the reference implementation with tools such as Certora or Slither, (iii) user studies evaluating how transfer caps influence purchasing behavior and perceived value, and (iv) cross-chain transfer accounting via bridge-integrated state proofs.

% ══════════════════════════════════════════════════════════════════════════════
%  9. CONCLUSION
% ══════════════════════════════════════════════════════════════════════════════

\section{Conclusion}
\label{sec:conclusion}

We introduce \textit{counted transfers} as a missing ownership primitive in NFT systems and realize this capability through ERC-7634, a minimal, backward-compatible extension to ERC-721 that bridges unrestricted transferability and permanent non-transferability. Our analysis shows that bounding ownership mobility fundamentally reshapes token dynamics by introducing a mobility premium, discouraging wash trading, and structurally limiting recursive collateralization in NFT-based protocols.

% ══════════════════════════════════════════════════════════════════════════════
%  REFERENCES
% ══════════════════════════════════════════════════════════════════════════════

\bibliographystyle{unsrt}
\bibliography{bib}

@misc{erc721,
  author       = {William Entriken and Dieter Shirley and Jacob Evans and Nastassia Sachs},
  title        = {{ERC-721}: Non-Fungible Token Standard},
  howpublished = {Ethereum Improvement Proposals},
  number       = {721},
  month        = jan,
  year         = {2018},
  url          = {https://eips.ethereum.org/EIPS/eip-721},
}

@misc{erc7634,
  author       = {Qin Wang and Saber Yu and Shiping Chen},
  title        = {{ERC-7634}: Limited Transfer Count {NFT}},
  howpublished = {Ethereum Improvement Proposals},
  number       = {7634},
  month        = feb,
  year         = {2024},
  url          = {https://eips.ethereum.org/EIPS/eip-7634},
}

@misc{erc5192,
  author       = {Tim Daubensch{\"u}tz and Anders},
  title        = {{ERC-5192}: Minimal Soulbound {NFTs}},
  howpublished = {Ethereum Improvement Proposals},
  number       = {5192},
  month        = jul,
  year         = {2022},
  url          = {https://eips.ethereum.org/EIPS/eip-5192},
}

@misc{erc5484,
  author       = {Buzz Cai},
  title        = {{ERC-5484}: Consensual Soulbound Tokens},
  howpublished = {Ethereum Improvement Proposals},
  number       = {5484},
  month        = aug,
  year         = {2022},
  url          = {https://eips.ethereum.org/EIPS/eip-5484},
}

@misc{erc4907,
  author       = {Anders and Lance and Shrug},
  title        = {{ERC-4907}: Rental {NFT}, an Extension of {EIP-721}},
  howpublished = {Ethereum Improvement Proposals},
  number       = {4907},
  month        = mar,
  year         = {2022},
  url          = {https://eips.ethereum.org/EIPS/eip-4907},
}

@misc{erc1155,
  author       = {Witek Radomski and Andrew Cooke and Philippe Castonguay and James Therien and Eric Binet and Ronan Sandford},
  title        = {{ERC-1155}: Multi Token Standard},
  howpublished = {Ethereum Improvement Proposals},
  number       = {1155},
  month        = jun,
  year         = {2018},
  url          = {https://eips.ethereum.org/EIPS/eip-1155},
}

@misc{erc2981,
  author       = {Zach Burks and James Morgan and Blaine Malone and James Seibel},
  title        = {{ERC-2981}: {NFT} Royalty Standard},
  howpublished = {Ethereum Improvement Proposals},
  number       = {2981},
  month        = sep,
  year         = {2020},
  url          = {https://eips.ethereum.org/EIPS/eip-2981},
}

@misc{erc5679,
  author       = {Zainan Victor Zhou},
  title        = {{ERC-5679}: Token Minting and Burning},
  howpublished = {Ethereum Improvement Proposals},
  number       = {5679},
  month        = sep,
  year         = {2022},
  url          = {https://eips.ethereum.org/EIPS/eip-5679},
}

@misc{erc6982,
  author       = {Francesco Sullo and Alexe Spataru},
  title        = {{ERC-6982}: Efficient Default Lockable Tokens},
  howpublished = {Ethereum Improvement Proposals},
  number       = {6982},
  month        = may,
  year         = {2023},
  url          = {https://eips.ethereum.org/EIPS/eip-6982},
}

@article{weyl2022decentralized,
  author       = {E. Glen Weyl and Puja Ohlhaver and Vitalik Buterin},
  title        = {Decentralized Society: Finding {Web3}'s Soul},
  journal      = {SSRN Electronic Journal},
  year         = {2022},
  month        = may,
  doi          = {10.2139/ssrn.4105763},
  url          = {https://doi.org/10.2139/ssrn.4105763},
}

@article{wang2022exploring,
  author       = {Qin Wang and Rujia Li and Qi Wang and Shiping Chen and Mark Ryan and Thomas Hardjono},
  title        = {Exploring Web3 From the View of Blockchain},
  journal      = {CoRR},
  volume       = {abs/2206.08821},
  year         = {2022},
  doi          = {10.48550/arXiv.2206.08821},
  url          = {https://arxiv.org/abs/2206.08821},
}

@inproceedings{von2022nft,
  author    = {Victor von Wachter and Johannes Rude Jensen and Ferdinand Regner and Omri Ross},
  title     = {{NFT} Wash Trading: Quantifying Suspicious Behaviour in {NFT} Markets},
  booktitle = {Financial Cryptography and Data Security. FC 2022 International Workshops},
  series    = {Lecture Notes in Computer Science},
  volume    = {13412},
  pages     = {299--311},
  year      = {2023},
  publisher = {Springer},
  doi       = {10.1007/978-3-031-32415-4_20},
  url       = {https://doi.org/10.1007/978-3-031-32415-4_20},
}

@inproceedings{das2022understanding,
author = {Das, Dipanjan and Bose, Priyanka and Ruaro, Nicola and Kruegel, Christopher and Vigna, Giovanni},
title = {Understanding Security Issues in the NFT Ecosystem},
year = {2022},
isbn = {9781450394505},
publisher = {Association for Computing Machinery},
address = {New York, NY, USA},
url = {https://doi.org/10.1145/3548606.3559342},
doi = {10.1145/3548606.3559342},
booktitle = {Proceedings of the 2022 ACM SIGSAC Conference on Computer and Communications Security},
pages = {667–681},
numpages = {15},
location = {Los Angeles, CA, USA},
series = {CCS '22}
}

@inproceedings{huang2024unveiling,
  author       = {Jintao Huang and Pengcheng Xia and Jiefeng Li and Kai Ma and Gareth Tyson and Xiapu Luo and Lei Wu and Yajin Zhou and Wei Cai and Haoyu Wang},
  title        = {Unveiling the Paradox of {NFT} Prosperity},
  booktitle    = {Proceedings of the ACM on Web Conference 2024},
  pages        = {167--177},
  year         = {2024},
  publisher    = {ACM},
  doi          = {10.1145/3589334.3645566},
  url          = {https://doi.org/10.1145/3589334.3645566},
}

@ARTICLE{wang2025manipulation,
  author={Wang, Bin and Gao, Yang and Wang, Wei},
  journal={IEEE Internet of Things Journal}, 
  title={Manipulation-Resilient Pricing for Non-Fungible Tokens}, 
  year={2025},
  volume={},
  number={},
  pages={1-1},
  doi={10.1109/JIOT.2025.3647747}}

@techreport{singh2010velocity,
  author       = {Manmohan Singh and James Aitken},
  title        = {The (Sizable) Role of Rehypothecation in the Shadow Banking System},
  institution  = {International Monetary Fund},
  type         = {IMF Working Paper},
  number       = {WP/10/172},
  year         = {2010},
  month        = jul,
  doi          = {10.5089/9781455201839.001},
  url          = {https://www.imf.org/en/Publications/WP/Issues/2016/12/31/The-Sizable-Role-of-Rehypothecation-in-the-Shadow-Banking-System-24075},
}

@techreport{nonfungible2024,
  author       = {{NonFungible.com}},
  title        = {Yearly {NFT} Market Report 2021},
  institution  = {NonFungible.com},
  year         = {2022},
  month        = mar,
  url          = {https://nonfungible.com/reports/2021/en/yearly-nft-market-report},
}

@article{qi2025understanding,
  author       = {Minfeng Qi and others},
  title        = {Understanding {NFT}s from {EIP} Standards},
  journal      = {CoRR},
  volume       = {abs/2508.07190},
  year         = {2025},
  doi          = {10.48550/arXiv.2508.07190},
  url          = {https://arxiv.org/abs/2508.07190},
}

@article{jiang2023decentralized,
  title={Decentralized finance (DeFi): A survey},
  author={Jiang, Erya and Qin, Bo and others},
  journal={arXiv preprint arXiv:2308.05282},
  year={2023}
}

@article{wang2021nft,
  author       = {Qin Wang and Rujia Li and Qi Wang and Shiping Chen},
  title        = {Non-Fungible Token ({NFT}): Overview, Evaluation, Opportunities and Challenges},
  journal      = {CoRR},
  volume       = {abs/2105.07447},
  year         = {2021},
  doi          = {10.48550/arXiv.2105.07447},
  url          = {https://arxiv.org/abs/2105.07447},
}

@inproceedings{sugino2025analysis,
  author       = {Takaya Sugino and Benjamin Kraner and James Angel and Shin'ichiro Matsuo and Rohil Paruchuri},
  title        = {An Analysis of Financial Stability Risk Propagation Through Leveraged Staking Activities},
  booktitle    = {International Conferences on Financial Cryptography and Data Security (FC) Workshops},
  series       = {Lecture Notes in Computer Science},
  volume       = {15754},
  pages        = {50--68},
  year         = {2025},
  publisher    = {Springer},
}

@article{wang2025understanding,
  title={Understanding daos: An empirical study on governance dynamics},
  author={Wang, Qin and Yu, Guangsheng and Sai, Yilin and Sun, Caijun and Nguyen, Lam Duc and Chen, Shiping},
  journal={IEEE Transactions on Computational Social Systems (TCSS)},
  year={2025},
  publisher={IEEE}
}

@article{yu2024toward,
  title={Toward web3 applications: Easing the access and transition},
  author={Yu, Guangsheng and Wang, Xu and Wang, Qin and Bi, Tingting and Dong, YiFei and Liu, Ren Ping and Georgalas, Nektarios and Reeves, Andrew},
  journal={IEEE Transactions on Computational Social Systems (TCSS)},
  volume={11},
  number={5},
  pages={6098--6111},
  year={2024},
  publisher={IEEE}
}

@article{wang2025understandingbrc,
  title={Understanding BRC-20: Hope or hype},
  author={Wang, Qin and Yu, Guangsheng and Chen, Shiping},
  journal={IEEE Transactions on Computational Social Systems (TCSS)},
  year={2025},
  publisher={IEEE}
}

@article{wang2024cryptocurrency,
  title={Cryptocurrency in the Aftermath: Unveiling the Impact of the SVB Collapse},
  author={Wang, Qin and Yu, Guangsheng and Chen, Shiping},
  journal={IEEE Transactions on Computational Social Systems (TCSS)},
  volume={11},
  number={5},
  pages={5839--5857},
  year={2024},
  publisher={IEEE}
}

\end{document}